\documentclass[apj, numberedappendix]{emulateapj}

\usepackage{graphicx} 
\usepackage{amsmath} 
\usepackage{amssymb}
\usepackage{exscale, relsize}
\usepackage{amscd}
\usepackage{natbib}

\begin{document}

\newcommand{\eg}{e.g.}
\newcommand{\ie}{\textit{i.e.}}
\newcommand{\viz}{\textit{viz.}}
\newcommand{\cf}{\textit{cf.}}

\newcommand{\reff}{R_\mathrm{eff}}
\newcommand{\zphot}{z_{\textrm{phot}}}
\newcommand{\zspec}{z_{\textrm{spec}}}
\newcommand{\zform}{z_{\textrm{form}}}
\newcommand{\delz}{{\cal D}_z}

\renewcommand{\sun}{$_{\odot}$}


\title{A Public, $K$--Selected, Optical--to--Near-Infrared Catalog \\
  of the Extended Chandra Deep Field South (ECDFS) \\ from the
  Multiwavelength Survey by Yale--Chile (MUSYC)}


\author{Edward N Taylor$^1$, 
  Marijn Franx$^1$, 
  Pieter G van Dokkum$^2$, 
  Ryan F Quadri$^1$, 
  Eric Gawiser$^3$, 
  Eric F Bell$^4$, \\ 
  L. Felipe Barrientos$^5$, 
  Guillermo A Blanc$^6$, 
  Francisco J Castander$^7$, 
  Maaike Damen$^1$, \\ 
  Violeta Gonzalez-Perez$^7$,
  Patrick B Hall$^8$, 
  David Herrera$^2$, 
  Hendrik Hildebrandt$^1$, \\ 
  Mariska Kriek$^9$, 
  Ivo Labb\'e$^{10}$, 
  Paulina Lira$^{11}$,  
  Jos\'e Maza$^{11}$, 
  Gregory Rudnick$^{12}$, \\
  Ezequiel Treister$^{13, 14}$, 
  C Megan Urry$^2$, 
  Jon P Willis$^{15}$, 
  Stijn Wuyts$^{16}$} 

\affil{$^1$ Sterrewacht Leiden, Leiden University, NL-2300 RA Leiden, Netherlands; ent@strw.leidenuniv.nl, 
$^2$ Department of Astronomy, Yale University, New Haven, CT 06520-8101, 
$^3$ Department of Physics and Astronomy, Rutgers University, Piscataway, NJ, 
$^4$ Max-Planck-Institut f\"ur Astronomie, D-69117 Heidelberg, Germany, 
$^5$ Departamento de Astronom\'ia, Pontificia Universidad Cat\'olica de
Chile, Santiago, Chile, 
$^6$ Department of Astronomy, University of Texas, Austin, TX, 
$^7$ Institu de Ci\`encies de l'Espai (IEEC/CSIC), Campus UAB, Facultat de
Ci\`encies, 08193 Bellaterra, Barcelona, Spain, 
$^8$ Department of Physics and Astronomy, York University, Toronto, Ontario M3J 1P3, Canada,
$^9$ H.N. Russell Fellow; Department of Astrophysical Sciences, Princeton University, Princeton, NJ 08544,
$^{10}$  Hubble Fellow; Carnegie Observatories, Pasadena, CA 91101,
$^{11}$  Departamento de Astronom\'ia, Universidad de Chile, Santiago, Chile,
$^{12}$ Goldberg Fellow; National Optical Astronomy Obervatories, Tucson, AZ,
$^{13}$ European Southern Observatory, Casilla 19001, Santiago 19, Chile,
$^{14}$ Chandra Fellow; Institute for Astronomy, 2680 Woodlawn Drive, University of Hawaii, Honolulu, HI 96822,
$^{15}$ Department of Physics and Astronomy, University of Victoria, Victoria, BC, Canada,
$^{16}$ W. M. Keck Postdoctoral Fellow, Harvard-Smithsonian Center for Astrophysics, Cambridge, MA 02138 }



\shorttitle{A $K$--Selected Catalog of the ECDFS from MUSYC}
\shortauthors{Taylor et al.}

\begin{abstract}
  We present a new, $K$--selected, optical--to--near infrared
  photometric catalog of the Extended Chandra Deep Field South
  (ECDFS), making it publicly available to the astronomical
  community.\footnote{Imaging and spectroscopy data and catalogs are
  freely available through the MUSYC Public Data Release webpage:
  http://www.astro.yale.edu/MUSYC/.}  The dataset is founded on
  publicly available imaging, supplemented by original $z'JK$ imaging
  data collected as part of the MUltiwavelength Survey by Yale--Chile
  (MUSYC).  The final photometric catalog consists of photometry
  derived from $UU_{38}BVRIz'JK$ imaging covering the full
  $\frac{1}{2} \times \frac{1}{2} ~ \square^\circ$ of the ECDFS, plus
  $H$ band photometry for approximately 80 \% of the field.  The $5
  \sigma$ flux limit for point--sources is $K{\mathrm{^{(AB)}_{tot}}}
  = 22.0$.  This is also the nominal completeness and reliability
  limit of the catalog: the empirical completeness for $21.75 < K <
  22.00$ is $\gtrsim 85$ \%.  We have verified the quality of the
  catalog through both internal consistency checks, and comparisons to
  other existing and publicly available catalogs.  As well as the
  photometric catalog, we also present catalogs of photometric
  redshifts and restframe photometry derived from the ten band
  photometry.  We have collected robust spectroscopic redshift
  determinations from published sources for 1966 galaxies in the
  catalog.  Based on these sources, we have achieved a (1$\sigma$)
  photometric redshift accuracy of $\Delta z/(1+z) = 0.036$, with an
  outlier fraction of 7.8 \%.  Most of these outliers are X-ray
  sources.  Finally, we describe and release a utility for
  interpolating restframe photometry from observed SEDs, dubbed
  InterRest\footnote{InterRest can be downloaded from
  http://www.strw.leidenuniv.nl/$\sim$ent/InterRest.  Documentation,
  including a complete walkthrough, is available from the same
  address.}.  Particularly in concert with the wealth of already
  publicly available data in the ECDFS, this new MUSYC catalog
  provides an excellent resource for studying the changing properties
  of the massive galaxy population at $z \lesssim 2$.
\end{abstract}

\section{Introduction}

Over the past decade, multi-band deep-field imaging surveys have
provided new opportunities to directly observe the changing properties
of the general, field galaxy population with lookback time.  These new
data, quantifying the star formation, stellar mass, and morphological
evolution among galaxies, have led to new and fundamental insights
into the physical processes that govern the formation and evolution of
galaxies.  These advances have been made possible not only by the
advent of a new generation of space-based and 8 m class telescopes,
but also the maturation of techniques for estimating redshifts and
intrinsic properties like stellar masses from observed SEDs.  These
two developments have made it possible not only to go deeper---pushing
to higher redshifts and probing further down the luminosity
function---but also to consider many more galaxies per unit observing
time.  This has made possible the construction of large,
representative, and statistically significant samples of galaxies
spanning a large proportion of cosmic time.

\begin{figure*} \centering
\includegraphics[width=16.5cm]{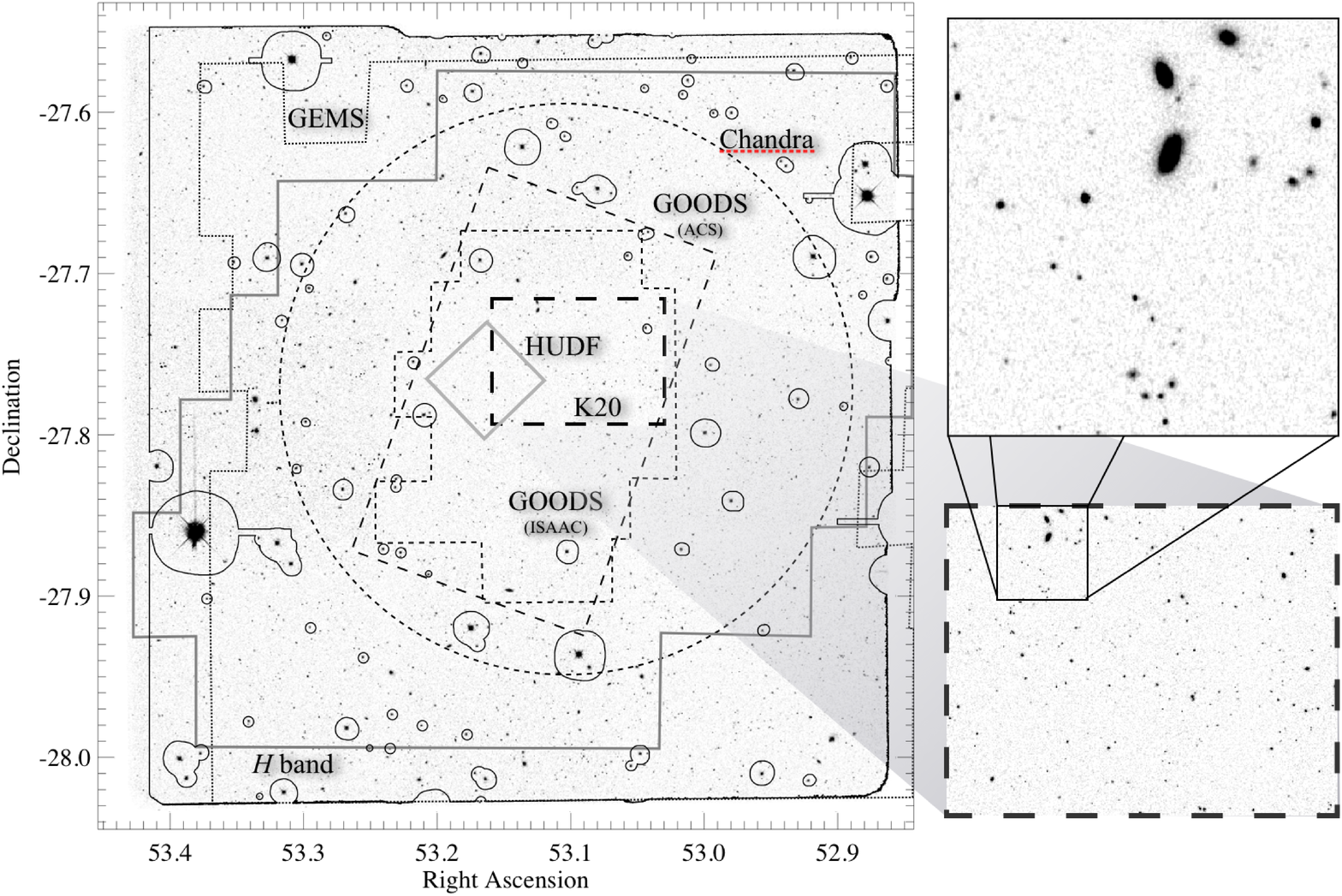}
\caption{MUSYC in the ECDFS.--- The greyscale image shows the new $K$
  band data.  The solid black contour shows the area with useful
  photometry in all of $UU_{38}BVRIz'JK$ in the MUSYC catalog.  (Areas
  badly affected by bright stars in the $z'$ band have been masked.)
  The catalog also includes $H$ photometry for $\sim 80 \%$ of the
  field (solid grey contour).  For comparison, we also show the area
  covered by several other important (E)CDFS surveys: GEMS
  \citep[][dotted lines]{RixEtAl}, the original Chandra CDFS
  \citep[][short-dashed circle]{GiacconiEtAl}, the GOODS
  \citep{Dickinson} HST ACS optical (light long-dashed rectangle) and
  ISAAC NIR (short-dashed region) imaging, the K20 survey
  \citep[][heavy long-dashed rectangle]{K20survey}, and the HUDF
  \citep[][grey solid diamond]{BeckwithEtAl}.  The FIREWORKS catalog
  \citep{WuytsEtAl} combines the GOODS ACS and ISAAC data with the
  $UU_{38}BRVIz'$ data described in this paper for the central GOODS
  ISAAC region.  SIMPLE (Damen et al., in prep.)\ will add very deep
  Spitzer IRAC imaging to the whole region shown here.  A medium band
  NIR survey is also underway using the NEWFIRM instrument
  \citep{newfirm}.  At right, we show a detail of the K20 survey area
  (below), and futher detail of an approximately $2'' \times 2''$ area
  (above). \label{fig:map}}
\end{figure*}


The Chandra Deep Field South \citep[CDFS;][]{GiacconiEtAl} is one of
the premier sites for deep field cosmological surveys (see Figure
\ref{fig:map}).  It is one of the most intensely studied region of the
sky, with observations stretching from the X-ray to the radio,
including ultraviolet, optical, infrared, and submillimeter imaging,
from space-based as well as the largest terrestrial observatories.  It
has also become traditional for surveys targeting the CDFS, to make
their data publicly available.  As a direct result of this commitment
to collaboration within the astronomical community, the wealth of data
available --- in terms of both volume and quality --- provide an
exceptional opportunity to quantify the evolution of the galaxy
population out to high redshift.

With this goal in mind, the key to gaining access to the $z \gtrsim 1$
universe is near infrared (NIR) data.  Most of the broad spectral
features (\eg\ the Balmer and 4000 \AA\ breaks) on which modern
SED--fitting algorithms rely are in the restframe optical; for $z
\gtrsim 1$, these features are redshifted beyond the observer's
optical window and into the NIR.  For this reason, we have combined
existing imaging of the Extended Chandra Deep Field South (ECDFS; see
Figure \ref{fig:map}) with new optical and NIR data taken as part of
the MUltiwavelength Survey by Yale--Chile (MUSYC).

The primary objective of MUSYC is to obtain deep optical imaging and
spectroscopy of four $\frac{1}{2} \times \frac{1}{2}~ \square^\circ$
Southern fields, providing parent catalogs for followup with ALMA.
Coupled with the optical ($UBVRIz'$) imaging program
\citep{GawiserEtAl}, there are two NIR components to the MUSYC
project: a deep component \citep[$K < 23.5$;][]{QuadriEtAl}, targeting
four $10 \times 10~ \square''$ regions within the MUSYC fields, and a
wide component \citep[$K < 22$;][this work]{BlancEtAl} covering three
of the $\frac{1}{2} \times \frac{1}{2}~ \square^\circ$ MUSYC fields in
their entirety.  These data are intended to allow, for example, the
restframe--UV selection of galaxies at $z \gtrsim 3$ using the Lyman
break technique \citep[\eg][]{SteidelEtAl}, the restframe--optical
selection of galaxies at $z \gtrsim 2$ using the Distant Red Galaxy
(DRG) criterion \citep{FranxEtAl}, and the color--selection of $z
\gtrsim 1.4$ galaxies using the $BzK$ criterion \citep{DaddiEtAl-BzK}.


\begin{table*} \begin{center}
\caption{Summary of the data comprising the MUSYC ECDFS catalog}
\begin{tabular*}{0.95\textwidth}{@{\extracolsep{\fill}}c c c c c c c c c c c c c}
\hline \hline
Band & $\lambda_0$ [\AA ]& $\Delta\lambda$ [\AA ] & $m_{\mathrm{Vega}}^{\mathrm{(AB)}}$ &
Int. Time [hr] & Area [$\square''$]& Eff. Seeing & $5 \sigma$ Depth & 
$N_\mathrm{cov}$ & $f_{5\sigma}$ &  $N_\mathrm{main}$ & $N_\mathrm{gals}$ & $N_\mathrm{stars}$ \\
(1) & (2) & (3) & (4) & (5) & (6) & (7) & (8) & (9) & (10) & (11) & (12) & (13) \\
\hline
$U$   & 3505 & 625 & $+1.01$  &  21.91  &  975  &  $1\farcs07$ & 26.5      & 15136   & 0.631  & 6213   & 6424  &  576 \\
$U_{38}$ & 3655 & 360 & $+0.82$   &  13.75  &  947  &  $1\farcs01$ & 26.0  & 14280   & 0.554  & 5505   & 5715  &  504 \\
$B$   & 4605 & 915 & $-0.12$ &  19.29  &  1012 &  $1\farcs03$ & 26.9       & 15153   & 0.852  & 8223   & 8322  &  880 \\
$V$   & 5383 & 895  & $-0.01$ &  29.06  &  1022 &  $0\farcs95$ & 26.6      & 15154   & 0.863  & 8370   & 8463  &  891 \\
$R$   & 6520 & 1600 & $+0.19$  &  24.35  &  1017 &  $0\farcs88$ & 26.3     & 15148   & 0.894  & 8647   & 8758  &  897 \\
$I$   & 8642 & 1500 & $+0.51$  &  9.60   &  977  &  $0\farcs98$ & 24.8     & 15128   & 0.826  & 8456   & 8545  &  897 \\
$z'$   & 9035 & 995 & $+0.54$ &  1.30   &  996  &  $1\farcs13$ & 24.0      & 13972   & 0.751  & 8043   & 8000  &  897 \\
$J$   & 12461 & 1620 & $+0.93$  &  1.33   &  906  & $\le 1\farcs49$ & 23.1 & 14580   & 0.683  & 7894   & 7859  &  896 \\ 
$H$   & 16534 & 2960 & $+1.40$  &  1.00   &  560 & $\le 1\farcs22$ & 23.1 & 10518   & 0.579  & 7005   & 6313  &  692 \\
$K$   & 21323 & 3310 & $+1.83$  &  1.00   &  906 & $\le 1\farcs05$ & 22.4  & 14355   & 0.695  & 8782   & 8911  &  897 \\
\hline \hline \label{tab:bands}
\end{tabular*} \end{center}
\tablecomments{For each band (Col.\ 1) that has gone into the MUSYC
  ECDFS catalog, we give the effective wavelength (Col.\ 2), the
  filter FWHM (Col.\ 3), and the apparent magnitude of Vega, in the AB
  system (\ie\ the conversion factor between the AB and Vega magnitude
  systems, Col.\ 4).  We also give the mean integration time (Col.\ 5)
  for each image, the effective imaging area (defined as the region
  receiving more than 75 \% of the nominal integration time, Col.\ 6),
  and the final effective seeing (FWHM, Col.\ 7).  The $5 \sigma$
  limiting depths given in Col.\ (8) are as measured in $2\farcs5$
  diameter apertures on the $1\farcs5$ FWHM PSF-matched images (see
  \textsection\ref{ch:psfmatching}; for a point source, these can be
  translated to total magnitudes by subtracting 0.45 mag.  Note that,
  whereas the optical data are taken in single pointings, the final
  NIR images are mosaics of many pointings.  Note that the central
  $\sim 10 \times 10~ \square''$ of the field received an extra three
  hours' integration time in the $H$ band; these data are
  approximately 0.3 mag deeper than the figure quoted above.  Col.\
  (9) gives the number of $K$ detections that useful coverage (\ie\ an
  effective weight, $w$, of 0.6 or greater) in each band ; Col.\ (10)
  gives the fraction of those objects that have $> 5\sigma$
  detections.  Both of these columns refer to the full catalog.
  Col.\ (11) gives the number of objects in the main science sample
  ($K_\mathrm{Tot} < 22$, $K$ S:N $>$ 5, $w_B > 0.6$, $w_{z'} > 0.6$,
  $w_K > 0.75$) with $>5 \sigma$ detections; Col.s (12) and (13) give
  the numbers of stars and galaxies separately (see
  \textsection\ref{ch:stargal}).}

\end{table*}

In the ECDFS, the broadband imaging data have been supplemented by a
narrow-band imaging survey, targeting Ly-$\alpha$ emitters at $z =
3.1$ \citep{GawiserEtAl2006,GronwallEtAl}, and a spectroscopic survey
\citep{TreisterEtAl} targeting Xray sources from the 250 ks ECDFS Xray
catalog \citep{LehmerEtAl, ViraniEtAl}.  Further, the Spitzer
IRAC/MUSYC Public Legacy in the ECDFS (SIMPLE; M Damen et al., in
prep.) project has obtained very deep IRAC imaging across the full
ECDFS.  There is also a deep medium band optical survey underway
(Cardamone et al., in prep.), and a planned medium band NIR survey
\citep{newfirm}.

This paper describes the MUSYC wide NIR--selected catalog of the ECDFS
(which we will from now on refer to as `the' MUSYC ECDFS catalog,
despite the existence of several separate MUSYC catalogs, as described
above), and makes it publicly available to the astronomical community.
A primary scientific goal of the wide NIR component of the survey is
to obtain statistically significant samples of massive galaxies at $z
\lesssim 2$.  In a companion paper \citep[][hereafter Paper
II]{TaylorEtAl}, we will use this dataset to quantify the $z \lesssim
2$ color and number density evolution of massive galaxies in general,
and in the relative number of red sequence galaxies in
particular. \looseness-1

The MUSYC ECDFS dataset is founded on existing and publicly available
imaging, supplemented by original optical ($z'$) and NIR ($JK$)
imaging.  Apart from the $JK$ imaging, all these data have been
described elsewhere.  Accordingly, the data reduction and calibration
of the new $JK$ imaging is a prime focus of this paper.  However, when
it comes to constructing panchromatic catalogs with legacy value from
existing datasets, the whole is truly more than the sum of parts:
ensuring both absolute and relative calibration accuracy is paramount.
We have invested substantial time and effort into checking all aspects
of our data and catalog, using both simulated datasets, and through
comparison to some of the many other existing (E)CDFS catalogs.

The structure of this paper is as follows: we describe the acquisition
and basic reduction of the MUSYC ECDFS broadband imaging dataset in
\textsection\ref{ch:data}.  The processes used to combine these data
into a mutually consistent whole are described in
\textsection\ref{ch:calibration}.  In \textsection\ref{ch:catalog}, we
describe the construction of the photometric catalog itself, including
checks on the completeness and reliability, and on our ability to
recover total fluxes.  We present external checks on the astrometric
and photometric calibration in \textsection\ref{ch:extras}.  After a
simple comparison of our catalog to other NIR--selected catalogs in
\textsection\ref{ch:nocounts}, we describe our basic analysis of the
multi-band photometry in \textsection\ref{ch:derived}, including
star/galaxy separation, and the derivation of photometric redshifts,
as well as the tests we have performed to validate our analysis.  In
\textsection\ref{ch:interrest}, we introduce InterRest; a new utility
for interpolating restframe fluxes.  This utility is also being made
public.  Additionally, in Appendix \ref{ch:speczs}, we describe a
compilation of 2213 robust spectroscopic redshift determinations for
objects in the MUSYC ECDFS catalog.

Throughout this work, all magnitudes are expressed in the AB system;
the only exception to this is \textsection\ref{ch:photcomps}, where
it will be convenient to adopt the Vega system.  Where necessary, we
assume the concordance cosmology; \viz\ $\Omega_\mathrm{m}=0.3$,
$\Omega_\Lambda = 0.7$, $\Omega_0 = 1.0$, and \mbox{$H_0=70$ km
s$^{-1}$ Mpc$^{-1}$}.  When discussing photometric redshifts, we will
characterise random errors in terms of the NMAD\footnote{Here, NMAD is
an abbreviation for the Normalized Median Absolute Deviation, and is
defined as $1.48 \times \mathrm{med} [x - \mathrm{med}(x)]$; the
normalization factor of 1.48 ensures that the NMAD of a Gaussian
distribution is equal to its standard deviation.}  of $\Delta
z/(1+z)$; we will abbreviate this quantity using the symbol
$\sigma_z$.


\section{Data} \label{ch:data}

This section describes the acquisition of the imaging data comprising
the MUSYC ECDFS dataset; the vital statistics of these data are given
in Table \ref{tab:bands}.  Of these data, only the $z'JK$ are
original; the WFI $UU_{38}BVRI$ imaging has been reduced and described
by \citet{HildebrandtEtAl}, and the SofI $H$ band data by
\citet{MoyEtAl} Further, the original $z'$ data have been reduced as
per \citet{GawiserEtAl} for the MUSYC optical ($BVR$---selected)
catalog.  We have therefore split this section between a summary of
the data that are described elsewhere (\textsection\ref{ch:existing}),
and a description of the new ISPI $JK$ imaging
(\textsection\ref{ch:ispidata}).  Note that what we refer to as the
$K$ band is really a `$K$ short' filter; we have dropped the subscript
for convenience.  For a complete description of the other datasets,
the reader is referred to the works cited above.


\subsection{Previously Described Data} \label{ch:existing}

\subsubsection{The WFI Data --- $UU_{38}BVRI$ Imaging from the ESO Archive}

\citet{HildebrandtEtAl} have collected all (up until December 2005)
archival $UU_{38}BRVI$\footnote{Two separate WFI $U$ filters have been
used.  The first, ESO\#877, which we refer to as the $U$ filter, is
slightly broader than a Broadhurst $U$ filter.  This filter is known
to have a red leak beyond 8000 \AA .  The second filter, ESO\#841,
which we refer to as $U_{38}$, is something like a narrow Johnson $U$
filter.  There is, unfortunately no clear convention for how to refer
to these filters; for instance, \citet{dps} refer to what we call the
$U$ and $U_{38}$ as $U'$ and $U$, respectively.} imaging data taken
using the Wide Field Imager \citep[WFI, $0\farcs238$
pix$^{-1}$;][]{BaadeEtAl98, BaadeEtAl99} on the ESO MPG 2.2 m telescope
for the four fields that make up the ESO Deep Public Survey
\citep[DPS;][]{dps}.  In addition the original DPS ECDFS data (DPS
field 2{\em c}), this combined dataset includes WFI commissioning
data, the data from the COMBO-17 survey \citep{WolfEtAl}, and
observations from seven other observing programs.
\citet{HildebrandtEtAl} have pooled and re-reduced these data using
the automated THELI pipeline described by \citet{ErbenEtAl} under the
moniker GaBoDS (Garching Bonn Deep Survey).  The final products are
publicly available through the ESO Science Archive
Facility.\footnote{http://archive.eso.org/cms/eso-data/data-packages/gabods-data-release-version-1.1-1/}
The final image quality of these images is $0\farcs9$---$1\farcs1$
FWHM.  \citet{HildebrandtEtAl} estimate that their basic calibration
is accurate to better than $\sim 0.05$ mag in absolute terms, and
that, based on color--color diagrams for stars, the relative or
cross-calibration between bands is accurate to $\lesssim 0.1$ mag for
all images.


\subsubsection{The Mosaic--II data---Original $z'$ Imaging}

We have supplemented the WFI optical data with original $z'$ band
imaging taken using Mosaic-II camera \citep[$0\farcs267$
pix$^{-1}$;][]{MullerEtAl} on the CTIO 4m Blanco telescope.  The data
acquisition strategy is the same as for the optical data in other
MUSYC fields \citep{GawiserEtAl}; the ECDFS data were taken in January
2005.  The final integration time was 78 minutes, with an effective
seeing of $1\farcs1$ FWHM, although we note that the PSF does have
broad, non-Gaussian `wings'.  The estimated uncertainty in the
photometric calibration is $< 0.03$ mag \citep{GawiserEtAl}.

\subsubsection{The SofI Data---$H$ Imaging Supporting the ESO DPS}

We include the $H$ band data described by \citet{MoyEtAl}, which was
taken to complement the original DPS WFI optical data and SofI NIR
data \citep{VandameEtAl, OlsenEtAl}.  This dataset covers
approximately 80 \% of the ECDFS, consisting of 32 separate $4\farcs9
\times 4\farcs9$ pointings, and were obtaining using SofI
\citep[$0\farcs288$ pix$^{-1}$;][]{MoorwoodEtAl} on the ESO NTT 3.6 m
telescope.  The data were taken as a series of dithered (or
`jittered') 1 min exposures, totaling 60 min per pointing; the central
four fields received an extra 3 hours' exposure time.  We received
these data (Pauline Barmby, priv.\ comm.)\ reduced as described by
\citet{MoyEtAl}; \ie, as 32 separate, unmosaicked fields.  The
effective seeing in each pointing varies from $0\farcs4$ to $0\farcs8$
FWHM.  \citet{MoyEtAl} found that their photometric zeropoint solution
varied by $\le 0.04$ mag over the course of a night; they offer this
as an upper limit on possible calibration errors.  Further, in
comparison to the Los Campanas Infrared Survey \citep[LCIRS;
][]{lcirsdata}, and the v0.5 (April 2002) release of the GOODS ISAAC
photometry, \citet{MoyEtAl} found their calibration to be 0.065 mag
brighter, and 0.014 mag fainter, respectively.

\subsection{The ISPI Data---Original $JK$ Imaging} \label{ch:ispidata}

The new MUSYC NIR imaging consists of two mosaics in the $J$ and $K$
bands, each made up of $3 \times 3$ pointings, and covering
approximately $950~\square '$.  The data were obtained using the
Infrared Sideport Imager (ISPI -- Probst et al.\ 2003; Van der Bliek
et al.\ 2004) on the CTIO Blanco 4m telescope.  ISPI uses a $2048
\times 2048$ pix HgCdTe HAWAII-2 detector, which covers approximately
$10\farcs5 \times 10\farcs5$ at a resolution of $\approx 0\farcs3$
pix$^{-1}$.  The aim was to obtain uniform $J$ and $K$ coverage of the
full $\frac{1}{2} \times \frac{1}{2}~\square^\circ$ of the ECDFS to
$\sim 80$ minutes and $\sim 60$ minutes, respectively; our target ($5
\sigma$, point source) limiting magnitudes were $J \approx 22.5$ and
$K \approx 22$.

The data were taken over the course of fifteen nights, in four
separate observing runs between January 2003 and February 2004.  In
order to account for the bright and variable NIR sky ($\sim 10000$
times brighter than a typical astronomical source of interest, varying
on many-minute timescales), the data were taken as a series of short,
dithered exposures.  A non-regular, semi-random dither pattern within
a $45''$ box was used for all but three sub-fields; these three
earliest pointings were dithered in regular, $\sim 10''$ steps.  An
exposure of $4 \times 15$s (\ie , 4 individual integrations of 15
seconds, coadded) was taken at each dither position in $K$; in $J$,
exposures were typically $1 \times 100$s. \looseness-1

Conditions varied considerably over the observing campaign, with
seeing ranging from $\lesssim 0\farcs7$ to $\gtrsim 1\farcs5$ FWHM.
All nine $K$ band pointings were observed under good conditions (
$\lesssim 1\farcs0$ FWHM).  However, observing condititions were
particularly bad for two of the nine $J$ pointings; the final
effective seeing of both the South and Southwest pointings are nearer
to $1\farcs5$ FWHM.

For each of the subfields comprising the MUSYC ISPI coverage of the
ECDFS, the data reduction pipeline is essentially the same as for the
other MUSYC NIR imaging, described by \citet{QuadriEtAl} and
\citet{BlancEtAl}, following the same basic strategy as, \eg ,
\citet{LabbeEtAl}.  The data reduction itself was performed using a
modified version of the IRAF package xdimsum.\footnote{IRAF is
distributed by the National Optical Astronomy Observatories, which are
operated by the Association of Universities for Research in Astronomy,
Inc., under cooperative agreement with the National Science
Foundation.  The xdimsum package is available from
\mbox{http://iraf.noao.edu/iraf/ftp/iraf/extern-v212/xdimsum020806}.}


\subsubsection{Dark Current and Flat Field Correction}

The ISPI detector has a non-negligible dark current.  To account for
this, nightly `dark flats' were constructed by mean combining
(typically) ten to twenty dark exposures with the appropriate exposure
times; these `dark flats' are then subtracted from each science
exposure.  These dark flats show consistent structure from night to
night, but vary somewhat in their actual levels.  Note that this
correction is done before flat-fielding and/or sky subtraction
\cite[see also][]{BlancEtAl}.

Flat field and gain/bias corrections (\ie , spatial variations in
detector sensitivity due to detector response, optic throughput, etc.)
were done using dome-flats, which were constructed either nightly or
bi-nightly.  These flats were constructed by taking a number of
exposures with or without a lamp lighting the dome screen.  Each
flatfield was constructed using approximately ten `lamp on' and `lamp
off' exposures, mean combined.  In order to remove background emission
from the `lamp on' image, we subtract away the `lamp off' image, to
leave only the light reflected from off of the dome screen \citep[see
also][]{QuadriEtAl}.  These flats are very stable night to night, with
some variation between different observing runs.


\subsubsection{Sky Subtraction and Image Combination} \label{ch:xdimsum}

Because the NIR sky is bright, non-uniform, and variable, a separate
sky or background image must be subtracted from each individual
science exposure.  The basic xdimsum package does this in a two-pass
procedure.  In the first pass, a background map is constructed for
each individual science image by median combining a sequence of
(typically) eight dithered but temporally continguous science
exposures: typically the four science images taken immediately before
and after the image in question.  In the construction of this
background image, a `sigma clipping' algorithm is used to identify
cosmic rays and/or bad pixels, which are then masked out.  The
resultant background image (which at this stage may be biased by the
presence of any astronomical sources) is then subtracted from the
science image to leave only astronomical signal.  The sky subtracted
images are then shifted to a common reference frame using the
positions of stars to refine the geometric solution, undoing the
dither, and then mean combined, again masking bad pixels/cosmic rays.
This combined image is used to identify astronomical sources, using a
simple thresholding algorithm.  This process is repeated in the second
`mask pass', with the difference that astronomical sources are now
masked when the background map is constructed.

Following \citet{QuadriEtAl}, we have made several modifications to
the basic xdimsum algorithm in order to improve the final image
quality.  We have constructed an initial bad pixel mask using the
flat-field images.  Further, each individual science exposure is
inspected by eye, and any `problem' exposures (especially those
showing telescope tracking problems or bad background subtraction) are
discarded; artifacts such as satellite trails and reflected light from
bright stars are masked by hand.  These masks are used in both the
first pass and mask pass.

Persistence is a problem for the ISPI detector: as a product of
detector memory, `echoes' of particularly bright objects linger for up
to eight exposures.  For this reason, we have also modified xdimsum to
create separate masks for such artifacts; these masks are used in the
mask pass.  Note that for the three subfields (including the Eastern
$K$ pointing) observed using a regular, stepped dither pattern, this
leads to holes in the coverage near bright objects: the `echoes' fall
repeatedly at certain positions relative to the source, corresponding
to the regular steps of the dither pattern.  At worst, coverage in
these holes is $\sim 25 \%$ of the nominal value.

Even after sky-subtraction, large-scale variations in the background
were apparent; these patterns were different and distinct for each of
the four quadrants of the images, corresponding to ISPI's four
amplifiers.  To remove these patterns, we have fit a 5th order
Legendre polynomial to each quadrant separately, using `sigma
clipping' to reduce the contribution of astronomical sources, and then
simply subtracted this away \citep[see also][]{BlancEtAl}.  This
subtraction is done immediately after xdimsum's normal
sky-subtraction.

In the final image combination stage, we adopt a weighting scheme
designed to optimize signal--to--noise for point sources \citep[see,
\eg,][]{GawiserEtAl, QuadriEtAl}.  At the end of this process, xdimsum
outputs a combined science image.  Additionally, xdimsum outputs an
exposure or weight map, and a map of the RMS in coadded pixels.  Note
that although this RMS map is not accurate in an absolute sense, it
does do an adequate job of mapping the spatial variation in the noise;
see \textsection\ref{ch:errors}.


\subsubsection{Additional Background Subtraction} \label{ch:backsub}

The sky subtraction done by xdimsum is imperfect; a number of large
scale optical artifacts (particularly reflections from bright stars
and `holes' around very bright objects) remain in the $K$ images as
output by xdimsum.  Using these images, in the object
detection/extraction phase, we were unable to find a combination of
SExtractor background estimation parameters (\viz\ BACK\_SIZE and
BACK\_FILTERSIZE) that was fine enough to map these and other
variations in the background but still coarse enough to avoid being
influenced by the biggest and brightest sources.  This led to
significant incompleteness where the background was low, and many
spurious sources where it was high.  We were therefore forced to
perform our own background subtraction, above and beyond that done by
xdimsum.

This basic idea was to use SExtractor `segmentation maps' associated
with the optical ($BVR$\footnote{Here, by $BVR$, we are referring to
the combined $B+V+R$ optical stack used for detection by
\citet{GawiserEtAl2006} in the construction of the MUSYC
optically-selected catalog of the ECDFS.}) and NIR ($K$) detection
images to mask real sources.  In particular, the much deeper $BVR$
stack includes many faint sources lying below the $K$ detection limit.
To avoid the contributions of low surface brightness galaxy `wings',
we convolved the combined ($BVR$+$K$) segmentation maps with a 15 pix
($4''$) boxcar filter to generate a `clear sky' mask.  Using this mask
to block flux from astronomical sources, we convolved the science
image with a 100 pix ($26\farcs7$) FWHM Gaussian kernel to generate a
new background map; this was then subtracted from the
xdimsum-generated science image.

Note that the background subtraction discussed above is important only
in terms of object {\em detection}; background subtraction for {\em
photometry} is discussed in \textsection\ref{ch:totalmags}.  While
this additional background subtraction step results in a considerably
flatter background across the detection image, it does not
significantly or systematically alter the measured fluxes of most
individual sources.

\subsubsection{Photometric Calibration}

Because not all pointings were observed under photometric conditions,
we have secondarily calibrated each NIR pointing separately with
reference to the 2MASS \citep{CutriEtAl, SkrutskieEtAl} Point Source
Catalog.\footnote{Available electronically via GATOR:
http://irsa.ipac.caltech.edu/applications/Gator/.} Taking steps to
exclude saturated, crowded, and extended sources, we matched ISPI
magnitudes measured in $16''$ diameter apertures to the 2MASS catalog
`default' magnitude (a $4''$ aperture flux, corrected to total
assuming a point--source profile).  For each subfield, the formal
errors on these zeropoint determinations are at the level of 1---2
percent.  The uncertainty is dominated by the 2MASS measurement
errors, and are highest for the central pointing where there are only
6---8 useful 2MASS--detected point sources.  For comparison, the
formal 2MASS estimates for the level of systematic calibration errors
is $\lesssim 0.02$ mag.


\section{Data Combination and Cross-Calibration} \label{ch:calibration}

This section is devoted to the combination and cross-calibration of
the distinct datasets described in the previous section into a
mutually consistent whole.  In \textsection\ref{ch:mosaicking}, we
describe the astrometric cross--calibration of each of the ten images,
including the mosaicking of the NIR data.  We describe and validate
our procedure for PSF-matching each band in
\textsection\ref{ch:psfmatching}.

\subsection{Astrometric Calibration and Mosaicking} \label{ch:mosaicking}

\begin{figure} \centering
\includegraphics[width=5.85cm]{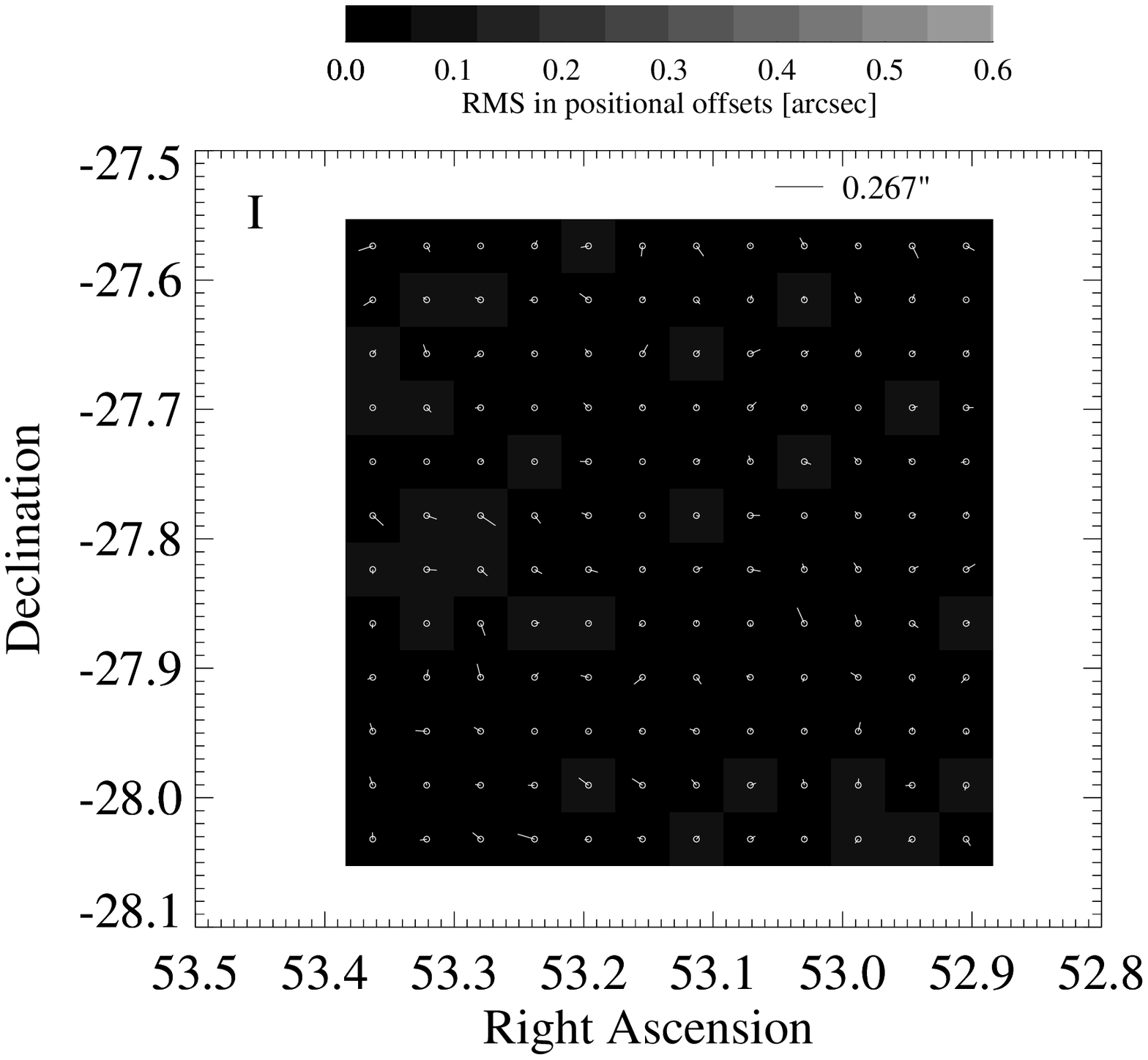}
\includegraphics[width=5.85cm]{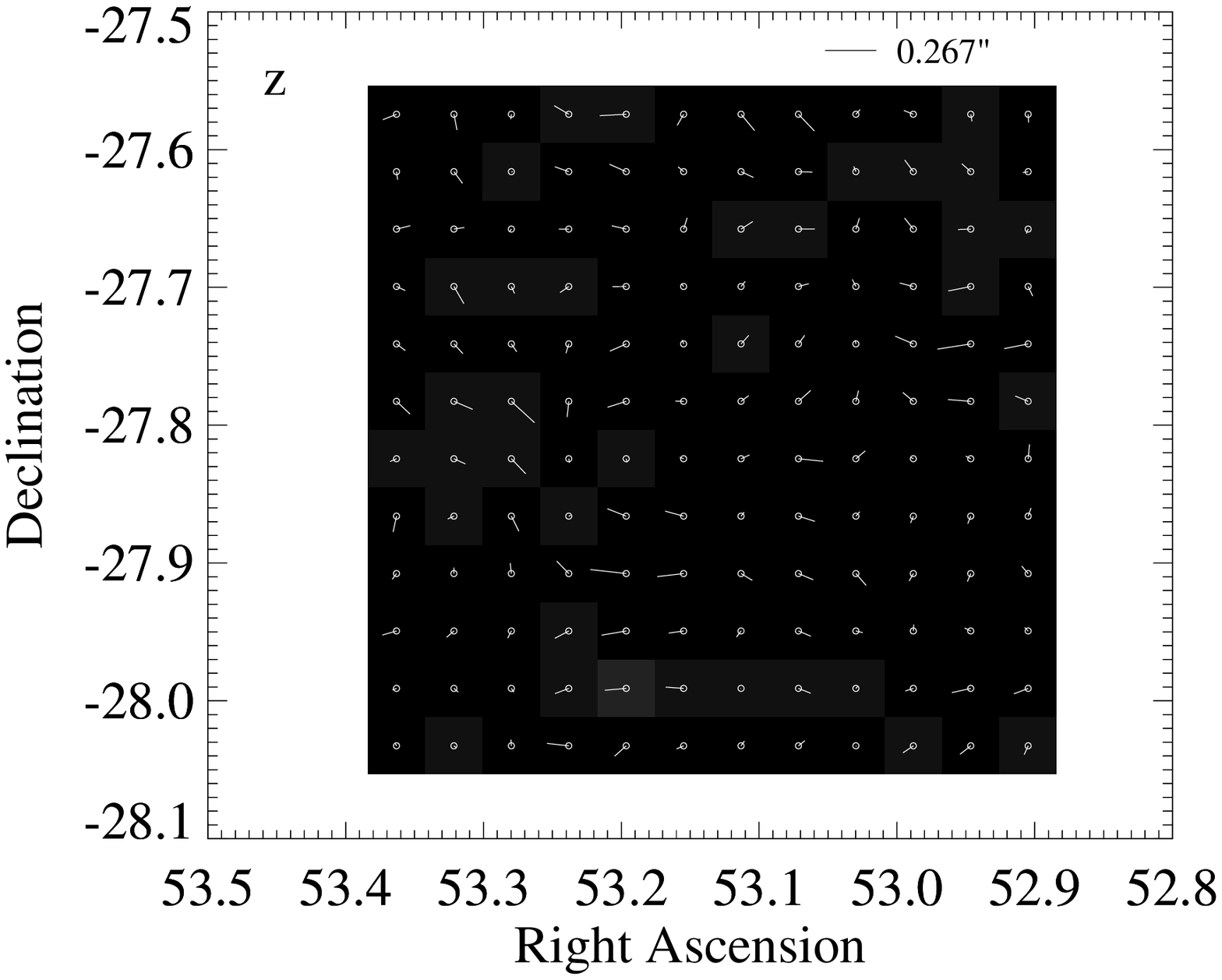}
\includegraphics[width=5.85cm]{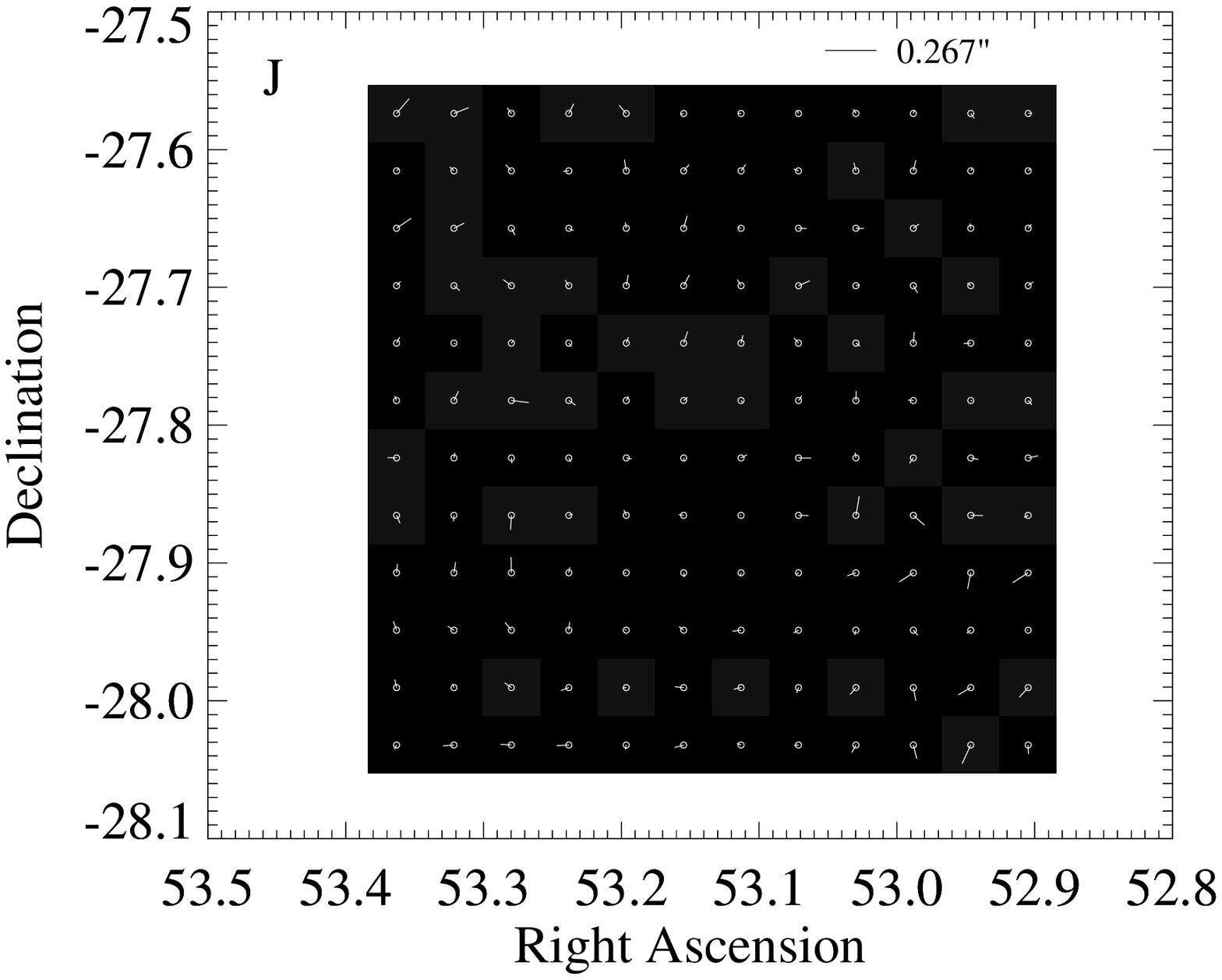}
\includegraphics[width=5.85cm]{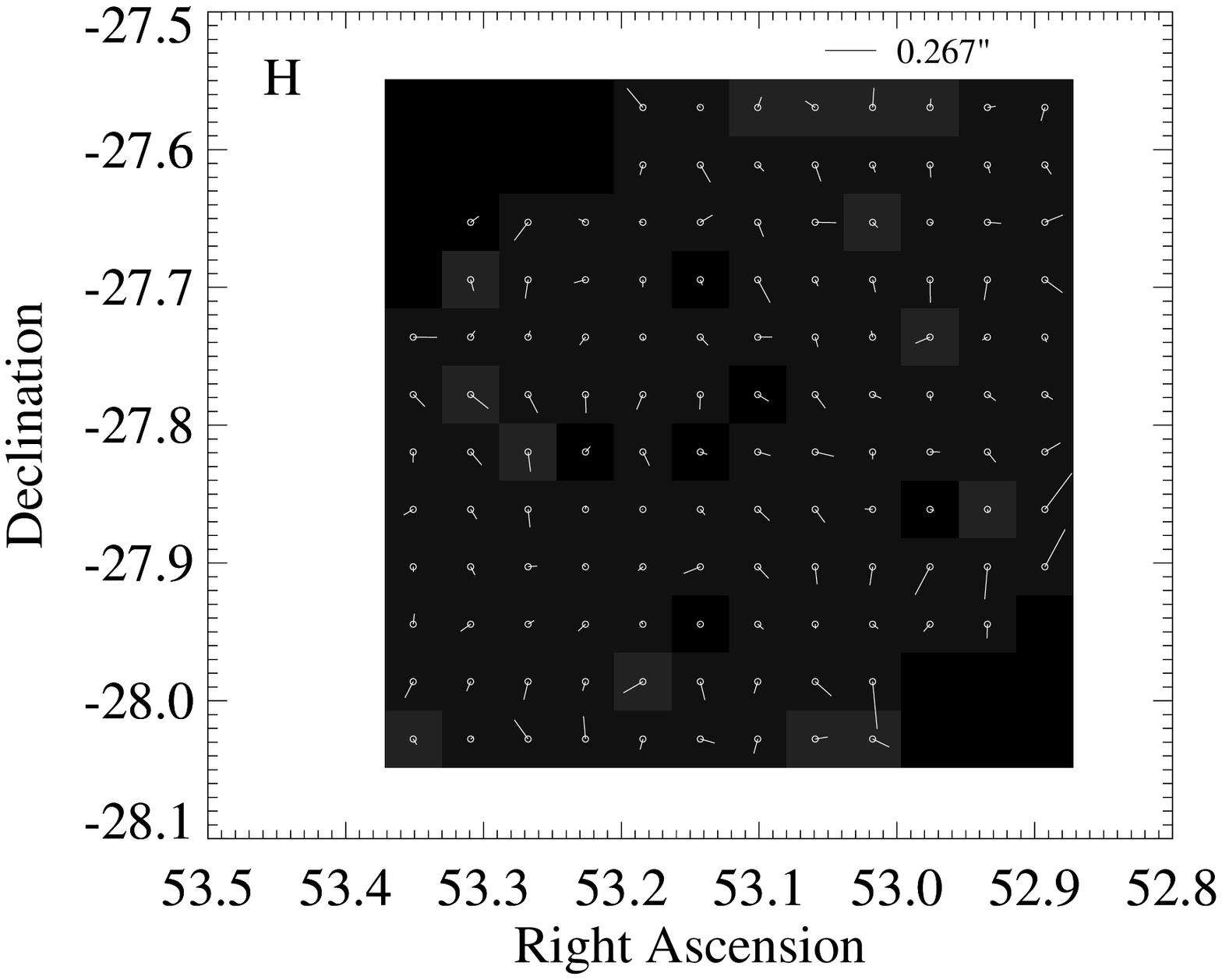}
\caption{The astrometric registration of the (from top to bottom)
  $Iz'JH$ images (obtaining using WFI, Mosaic-II, ISPI, and SofI,
  respectively), relative to the K detection image ---In each panel,
  vectors give the biweight mean positional offset between the two
  images in $2\farcm5 \times 2\farcm5$ cells, based on all $K < 22$
  sources; the greyscale gives the biweight variance.  Systematic
  astrometric shears in individual images images are typically much
  less than a pixel.\label{fig:astrometry}}
\end{figure}

To facilitate multi-band photometry, each of the final science images
is transformed to a common astrometric reference frame: a North-up
tangential plane projection, with a scale of $0\farcs267$ pix$^{-1}$.
This chosen reference frame corresponds to the stacked $BVR$ image
used as the detection image for the optically-selected MUSYC ECDFS
catalog \citep[see][]{GawiserEtAl,GawiserEtAl2006}, based on an early
reduction of the WFI data.

Whereas WFI and Mosaic-II are both able to cover the entire ECDFS in a
single pointing, the SofI and ISPI coverage consists of 32 and nine
subfields, respectively.  For these bands, each individual subfield
was astrometrically matched to the $BVR$ reference image using
standard IRAF/PyRAF tasks.  For the ISPI data, each subfield is then
combined, weighted by S:N on a per pixel basis, in order to create the
final mosaicked science image.  (Note that individual subfields are
also `PSF-matched' before mosaicking -- see
\textsection\ref{ch:psfmatching}.)

One severe complication in this process is that exposure/weight maps
were not available for the SofI imaging.  We have worked around this
problem by constructing mock exposure maps based on estimates of the
per pixel RMS in each science image.  Specifically, we calculate the
biweight scatter in rows and columns: $\sigma_B(x)$ and
$\sigma_B(y)$.  The effective weight for the pixel ($x$, $y$) is then
estimated as $[\sigma_B(x)\sigma_B(y)]^{-2}$.  The map for each
subfield is normalized so that the median weight is 1 for those
pointings that received 1 hour's integration, and 4 for the four
central pointings.

In line with \citet{QuadriEtAl}, we found it necessary to fit a high
order surface (\viz , a 6th order Legendre polynomial, including $x$
and $y$ cross terms) to account for the distortions in the ISPI focal
plane.  For the SofI data, a 2nd order surface was sufficient,
although we did find it necessary to revise the initial astrometric
calibration by \citet{MoyEtAl}.

As an indication of the relative astrometric accuracy across the whole
dataset, Figure \ref{fig:astrometry} illustrates the difference
between the positions of all $K < 22$ sources measured from the K
band, and those measured in each of the $Rz'JH$ bands (observed using,
in order, WFI, Mosaic-II, ISPI, and SofI).  Systematic `shears'
between bands are typically much less than a pixel.  Comparing
positions measured from the registered $R$ and $K$ band images,
averaged across the entire field, the mean positional offset is
$0\farcs15$ (0.56 pix).  Looking only at the $x$/$y$ offsets, we find
the biweight mean and variance to be $0\farcs03$ (0.11 pix) and
$0\farcs3$ (1.1 pix), respectively.


\subsection{PSF Matching} \label{ch:psfmatching}

The basic challenge of multi-band photometry is accounting for
different seeing in different bands, in order to ensure that the same
fraction of light is counted in each band for each object.  We have
done this by matching the PSFs in each separate pointing to that with
the broadest PSF.  Of all images, the South-Western $J$ pointing has
the broadest PSF: $1\farcs5$ FWHM.  This sets the limiting seeing for
the multiband SED photometry.  Among the $K$ pointings, however, the
worst seeing is $1\farcs0$ FWHM; this sets the limiting seeing for
object detection, and the measurement of total $K$ magnitudes (see
\textsection\ref{ch:detection} and \textsection\ref{ch:totalmags}).
We have therefore created eleven separate science images: one
$1\farcs5$ FWHM image for each of the ten bands to use for SED
photometry, plus a $1\farcs0$ FWHM $K$ image for object detection and
the measurement of total $K$ fluxes.

The PSF-matching procedure is as follows: for each pointing, we take a
list of SED--classified stars from the COMBO-17 catalog; these
objects are then used to construct an empirical model of the PSF in
that image, using an iterative scheme to discard low
signal--to--noise, extended, or confused sources.  Our results do not
change if we begin with $Bz'K$ selected stars, or GEMS point sources.
We then use the IRAF/PyRAF task lucy (an implementation of the
Lucy-Richardson deconvolution algorithm, and part of the STSDAS
package\footnote{STSDAS is a product of the Space Telescope Science
Institute, which is operated by AURA for NASA.}) to determine the
convolution kernel required to `degrade' each subfield to the limiting
effective seeing.  Finally, the convolution is done using standard
tasks.  Note that each of the NIR subfields is treated individually,
prior to mosaicking.

\begin{figure} \centering 
\includegraphics[width=8.6cm]{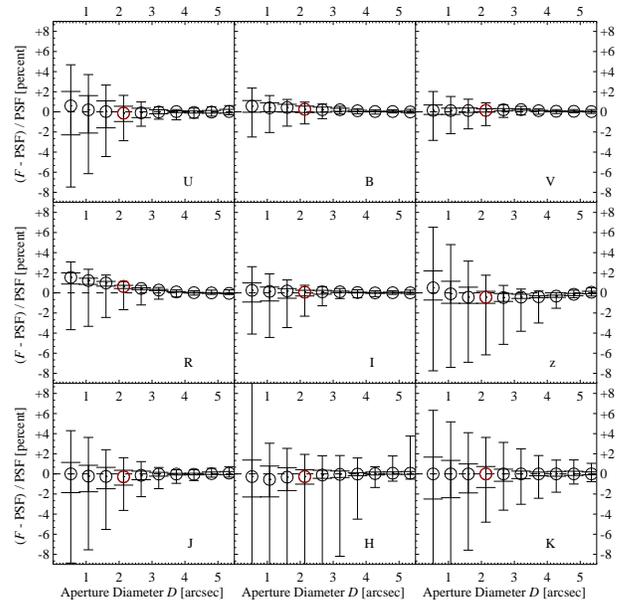}
\caption{Relative deviations in the curves of growth for
  point--sources in each of nine bands, from four different
  instruments, after PSF matching ($1\farcs5$ FWHM)---Each panel
  shows the relative differences between the normalized growth curves
  of bright, unsaturated, isolated point sources, plotted as a
  function of aperture diameter. Circles show the median of all growth
  curves in each band; large and small error bars show the 33/67 and
  5/95 percentiles, respectively.  The growth curves in different
  bands are all normalized with respect to the $K$ band median; the
  systematic errors in the $K$ panel are thus zero by construction.
  For our smallest apertures ($2\farcs5$), systematic offsets due to
  imperfect PSF matching are at worst 0.006 mag; random errors, due
  to, for example, spatial variation of the PSF, are $\lesssim 0.03$
  mag.
  \label{fig:psfmatching}}
\end{figure}

In order to quantify the random and systematic errors resulting from
imperfect PSF matching, Figure \ref{fig:psfmatching} shows the
relative difference between the curves of growth of individual point
sources across nine of our ten bands, after matching to the target
$1\farcs5$ FWHM PSF.  In this Figure, we compare the growth curves of
many bright, unsaturated, isolated point sources as a function of
aperture diameter; specifically, we plot the relative difference
between the normalized growth curves in each band, compared to the
median $K$ band growth curve.  Within each panel, the circles
represent the median growth curve in each band (zero for the $K$ band
by construction), and the large and small error bars represent the
33/67 and 5/95 percentiles, respectively.

After PSF matching, there are signs of spatial variations in the FWHM
of the $J$ and $K$ PSFs at the few percent level, particularly towards
the edges of each pointing.  But since the scatter in these plots
represents both real spatial deviations in the PSF, as well as
normalization errors, these results can thus can be taken as an upper
limit on the random PSF--related photometric errors.  Looking at the
$z'$-band panel, it is possible that the broad $z'$ band PSF wings are
important at the $\lesssim 0.005$ mag level for
$2\farcs5$---$5\farcs0$.  Note that the smallest apertures we use are
$2\farcs5$ in diameter --- for these apertures, random errors due to
imperfect PSF matching are typically $\lesssim 0.03$ mag, and
systematic errors are at worst $0.006$ mag.


\section{Detection, Completeness, Photometry, and Photometric Errors}
\label{ch:catalog}

In this section, we describe our scheme for building our multi-colour
catalog of the ECDFS; a summary of the contents of the final
photometric catalog is given in Table \ref{tab:photcat}.  We rely on
SExtractor \citep{BertinArnouts} for both source detection and
photometry; in \textsection\ref{ch:detection} we describe our use of
SExtractor, and we quantify catalog completeness and reliability in
\textsection\ref{ch:completeness}.  There are two separate components
to the reported photometry for each object: the total $K$ flux, which
is discussed in \textsection\ref{ch:totalmags} and
\textsection\ref{ch:magsvalid}, and the ten band SED, which is
discussed in \textsection\ref{ch:seds}.  Finally, in
\textsection\ref{ch:errors}, we describe the process by which we have
quantified the photometric measurement uncertainties.

\begin{table*} \centering
\caption{Summary of the Contents of the Photometric Catalog}
\label{tab:photcat}
\begin{tabular*}{0.95\textwidth}{@{\extracolsep{\fill}}l l p{12 cm}}
\hline\hline
\\
Column No. & Column Title & Description \\
\\
\hline\hline
\\
1 & id & Object identifier, beginning from 1 \\

2, 3 & ra, dec & Right ascension and declination (J2000), expressed in
decimal degrees \\

4 & field & An internal MUSYC field identifier (ECDFS=8) \\

5, 6 & x, y & Center of light position, expressed in pixels \\

7 & ap\_col & Effective diameter (\ie\ $\sqrt{4\pi A}$, where $A$ is
the aperture area), in arcsec; we use the larger of SExtractor's ISO
aperture and a $2\farcs5$ diameter aperture to measure colors (see
\textsection\ref{ch:seds}) \\

8---27 & U\_colf, U\_colfe, etc. & Observed flux,$^a$ with the
associated measurement uncertainty, in each of the $UU_{38}BVRIz'JHK$
bands, as measured in the 'color' aperture \\

28  & ap\_tot & Effective diameter of the AUTO aperture, on which the total $K$ flux measurement is based \\

29, 30 & K\_totf, K\_totfe & Total $K$ flux---based on SExtractor's
AUTO measurement, with corrections applied for missed flux and
background over-subtraction (see \textsection\ref{ch:totalmags})---and
the associated measurement uncertainty, which accounts for correlated
noise, random background subtraction errors, spatial variations in the
noise, Poisson shot noise, etc.\ (see \textsection\ref{ch:errors}) \\

31, 32 & K\_4arcsecf, K\_4arcsecfe & $K$ flux, as measured in a $4''$ aperture, with the associated measurement uncertainty \\ 

33, 34 & K\_autof, K\_autofe & $K$ flux within SExtractor's AUTO aperture, with the associated measurement uncertainty \\ 

35---37 & Kr50, Keps, Kposang & Morphological parameters from
SExtractor, measured from the $1\farcs0$ FWHM $K$ image; \viz , the
half-light radius (where the `total' light here is the AUTO flux),
ellipticity, and position angle \looseness-1 \\

38---47 & Uw, etc. & Relative weight in each of the $UU_{38}BVRIz'JHK$
bands.$^{b}$ \\

48 & id\_sex & The original SExtractor identifier,$^c$ for use with
the SExtractor generated segmentation map \\

49, 50 & f\_deblend1 f\_deblend2 & Deblending flags from SExtractor,
indicating whether an object has been deblended, and whether that
object's photometry is significantly affected by a near neighbor,
respectively \\

51 & star\_flag & A flag indicating whether an object's $Bz'K$ color
suggests its being a star (see \textsection\ref{ch:stargal}) \\

52---54 & z\_spec, qf\_spec, spec\_class & Spectroscopic redshift
determination, if available, along with the associated quality flag
and spectral classification, if given. \\

55, 56 & source, nsources & A code indicating the source of the
spectroscopic redshift, and the number of agreeing determinations \\

57, 58 & qz\_spec, spec\_flag & A figure of merit, derived from the MUSYC
photometry, for the spectroscopic redshift determination (see Appendix
\ref{ch:speczs}), and a binary flag indicating whether the
spectroscopic redshift is considered `secure'\\
\\
\hline \hline
\end{tabular*}
\tablecomments{$^a$ All fluxes are given in such a way that they can
be transformed to AB magnitudes using a zeropoint of 25; in other
words, fluxes are given in units of 0.363 mJy.  $^b$ For all but the
$z'$ and $H$ bands, this is essentially the exposure time, normalized
by the nominal values given in Table \ref{tab:bands}.  For the $H$
band, this value is derived from the mock exposure map described in
\textsection\ref{ch:mosaicking}; the $z'$ band, this is a binary flag
indicating whether the $z'$ photometry is significantly affected by
light from a nearby bright star. $^c$ Recall that we have excised all
detections with an effective exposure time of less than 12 minutes
from the catalog output by SExtractor.}
\end{table*}

\subsection{Detection} \label{ch:detection}

Source detection and photometry for each band was performed using
SExtractor in dual image mode; that is, using one image for detection,
and then performing photometry on a second `measurement' image.  In
all cases, the $1\farcs0$ FWHM $K$ band mosaic (see \textsection
\ref{ch:psfmatching}) was used as the detection image; since flexible
apertures are always derived from the detection image, this assures
that the same apertures are used for all measurements in all bands.

As a standard part of the SExtractor algorithm, the detection image is
convolved with a `filter' function that approximates the PSF; we use a
4 pix ($\sim 1\farcs0$) FWHM Gaussian filter.  We adopt an absolute
detection threshold equivalent to 23.50 mag / $\square''$ in the
filtered image, requiring 5 or more contiguous pixels for a detection.
Since we have performed our own background subtraction for the NIR
images (see \textsection\ref{ch:backsub}), we do not ask SExtractor to
perform any additional background subtraction in the detection phase.
For object deblending, we set the parameters DEBLEND\_NTHRESH and
DEBLEND\_MINCONT to 64 and 0.001, respectively.  These settings have
been chosen by comparing the deblended segmentation map for the $K$
detection image to the optical $BVR$ detection stack, which has a
considerably smaller PSF.

Near the edges of the observed region, where coverage is low, we get a
large number of spurious sources.  We have therefore gone through the
catalog produced by SExtractor, and culled all objects where the $K$
effective weight, $w_K$, is less than 0.2 (equivalent to $\lesssim 12$
minutes per pointing).  This makes the effective imaging area $953 ~
\square ''$.  Further, we find that a large number of spurious sources
are detected where there are `holes' in the coverage map (a product of
the regular dither pattern used for the earliest eastern and
northeastern tiles; see \textsection\ref{ch:mosaicking}.)  To avoid
these spurious detections, for scientific analysis we will consider
only those detections with an $w_k > 0.75$ (equivalent to $\sim 45$
minutes per pointing) or greater.\footnote{In other words, the
catalog is based on the area that received the equivalent of
$\gtrsim 12$ min integration, but our scientific analysis is based on
those objects that received $\gtrsim 45$ min integration.  While
objects with $0.2 < w_K < 0.75$ are given in the catalog, we do not
include them in our main science sample, because of the poorer
completeness and reliability among these objects.}  This selection
reduces the effective area of the catalog to $887~ \square ''$.


\begin{figure}[b] \centering
\includegraphics[width=8.8cm]{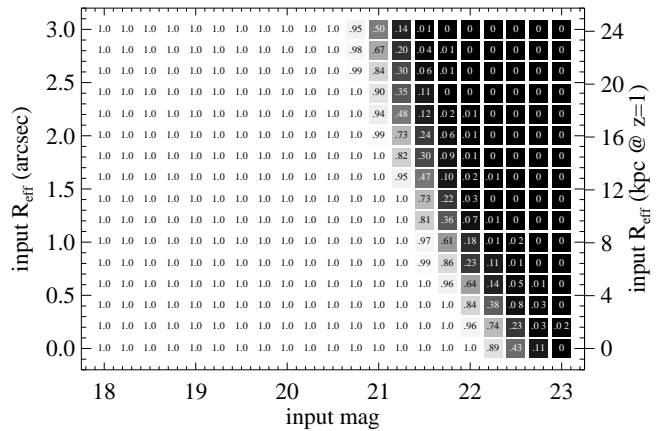}
\caption{Completeness for synthetic $R^{1/4}$--law sources added to
  the $1\farcs0$ FWHM $K$ detection mosaic---At each point, we give
  the fraction of synthetic sources (characterized by their total $K$
  magnitude and effective or halflight radius, $\reff$) that we are
  able to recover, using identical processes as for `live' detection.
  Each synthetic source has been isolated from any other real or
  synthetic source by at least 50 pix to prevent confusion.  While we
  are 100 \% complete for point sources (\ie\ $\reff = 0$) at our
  nominal limit of $K=22$, the completeness drops rather rapidly for
  larger radii/lower surface brightnesses.  At fixed $K$ and $\reff$,
  both surface brightness and completeness are strong functions of the
  profile shape; we present these results for $R^{1/4}$ sources as
  loose lower limits on the true completeness.
\label{fig:completeness} }
\end{figure}

\subsection{Completeness and Reliability} \label{ch:completeness}

In order to estimate the catalog completeness, we have added a very
large number of simulated sources into the $1\farcs0$ FWHM detection
image, and checked which are recovered by SExtractor, using the same
settings as `live' detection.  The completeness is then just the
fraction of inputed sources which are recovered, as a function of
source size and brightness.  We adopted a de Vaucouleurs
($R^{1/4}$--law) profile for all simulated sources, each with a
half-light radius, $\reff$, between $0''$ (\ie\ a point source) and
$3''$, an ellipticity of 0.6, and total apparent $K$ magnitude in the
range 18---23 mag.  We truncate each object's profile at 8 $\reff$.
No more than 750 artificial galaxies were added at any given time,
corresponding to 3--5 \% increase in the number of detected sources.
Simulated sources were placed at least $13\farcs35$ (50 pix) away from
any other detected or simulated source; these completeness estimates
therefore do not account for confusion.

The results of this exercise are shown in Figure
\ref{fig:completeness}, which plots the completeness as a function of
size and brightness.  For point sources, we are 50 \%, 90 \%, and 95
\% complete for K = 22.4, 22.2, and 22.1 mag, respectively.  At a
fixed total magnitude, the completeness drops for larger, low surface
brightness objects.  At $K = 22$, the nominal completeness limit of
the catalog, we are in fact only 84 \% complete for $\reff =
0\farcs4$, assuming an $R^{1/4}$ profile.  Note that we detect quite a
few objects that `really' lie below our formal (surface brightness)
detection limit: just as noise troughs can `hide' galaxies, noise
peaks can help push objects that would not otherwise be detected over
the detection threshold.  (See also \textsection\ref{ch:totalmags}.)
\looseness-1

Note that the above test explicitly avoids incompleteness due to
source confusion.  If we repeat the above test without avoiding known
sources, we find that where completeness is low, confusion actually
increases the completeness by a factor of a few, with faint sources
hiding in the skirts of brighter ones \citep[see also][]{Berta}.
However, the flux measurements for these objects are naturally
dominated by their neighbours; in this sense, it is arguable as to
whether the synthetic object is actually being `detected'.  Where
completeness is high ($K \lesssim 20.5$), confusion reduces
completeness by a few percent, but again, the exact amount is
sensitive to the position and flux agreement required to define a
successful detection.  From these tests, it seems that $\lesssim 2$ \%
of sources are affected by confusion due to chance alignments with
foreground/background galaxies (\cf\ gravitational associations).  For
comparison, based on the SExtractor segmentation map, $K$--detected
objects cover 2.34 \% of the field.

\begin{figure} \centering
\includegraphics[width=8.8cm]{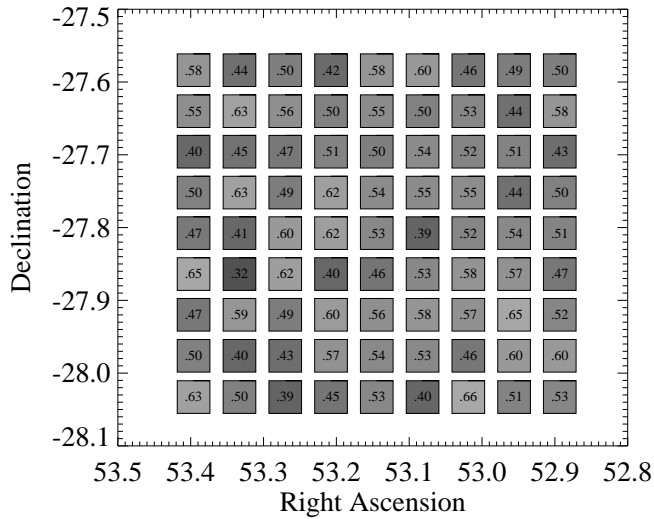}
\caption{Spatial Variation in the completeness of the MUSYC ECDFS
  catalogs--- Completeness for synthetic $K = 22.4$ point sources
  randomly added to the K detection image.  Completeness is slightly
  lower around the very bright star towards the eastern edge of the
  field, but is otherwise reasonably uniform. \label{fig:fieldcomp}}
\end{figure}

We have also done a similar test to investigate any variations in
completeness across the field.  We placed 5000 point sources with K =
22.4 --- our 50 \% completeness limit for point sources --- across the
field, each isolated by at least $26\farcs7$ (100 pix).  The results
are shown in Figure \ref{fig:fieldcomp}.  Although it is perhaps
slightly lower for the noisier east and northeast pointings, the
completeness is indeed quite uniform across the full field.

\begin{figure} \centering
\includegraphics[width=8.8cm]{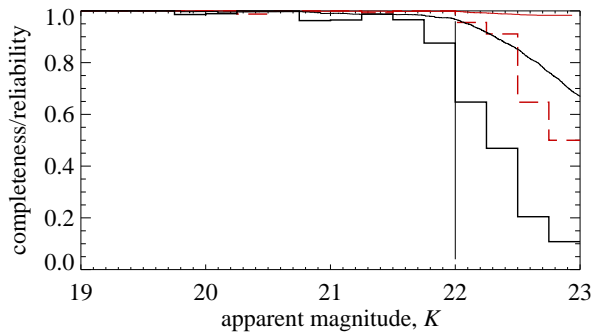}
\caption{Completeness and reliability of the MUSYC ECDFS catalog in
  comparison to the FIREWORKS catalog \citep{WuytsEtAl} of the
  GOODS-CDFS data---We show the fraction of FIREWORKS sources that are
  detected by MUSYC (\ie , the completeness of the MUSYC catalog; {\em
  solid black histograms}), and the fraction of MUSYC sources that are
  confirmed by FIREWORKS (\ie , the reliability of the MUSYC catalog;
  {\em dashed red histograms}), in bins of total apparent $K$
  magnitude.  For the $21.75 < K < 22.00$ bin, we are more than 85 \%
  complete, with essentially all detections confirmed by the (much
  deeper) GOODS data.  Cumulatively, to $K < 22$, the MUSYC catalog is
  97 \% complete ({\em black curve}), and better than 99 \% reliable
  ({\em red curve}). \label{fig:goodscomp} }
\end{figure}

Finally, we can obtain empirical measures of both completeness and
reliability by comparing our catalog to the much deeper $K$--selected
FIREWORKS catalog of the GOODS-CDFS region \citep{WuytsEtAl} The
results of this exercise are shown in Figure \ref{fig:goodscomp}.
Here, the completeness is just the fraction of FIREWORKS sources which
also appear in the MUSYC catalog; similarly, the reliability is the
fraction of MUSYC sources which do not appear in the FIREWORKS
catalog.  For the $21.8 < K < 22.0$ bin, the MUSYC catalog is 87.5 \%
complete, and 97 \% reliable.  For $K < 22$, the overall completeness
and reliability are 97 \% and 99 \%, respectively.

Since the GOODS-ISAAC data are so much deeper, the high completeness
at K $\sim 22$ implies that K $\sim 22$, $R_e \gtrsim 0\farcs5$
objects make up at most a small fraction of the FIREWORKS catalog.
This might imply that our catalog is primarily flux, rather than
surface brightness, limited.  It must also be remembered, however,
that the main motivation for large area surveys like MUSYC is to find
the rare objects that may be missed in smaller area surveys like
GOODS.


\subsection{Total Fluxes --- Method} \label{ch:totalmags}

We measure total fluxes in the $1\farcs0$ FWHM $K$ band mosaic, using
SExtractor's AUTO measurement, which uses a flexible elliptical
aperture whose size ultimately depends on the distribution of light in
`detection' pixels (\ie , an isophotal region).  We do specify a
minimum AUTO aperture size (using the parameter PHOT\_AUTOAPERS) of
$2\farcs5$, although in practice this limit is almost never reached
for sources with $K < 22$.  The $2\farcs5$ limit has been chosen to be
small enough to ensure high signal-to-noise for faint point sources,
while still avoiding any significant aperture matching effects (see
both \textsection\ref{ch:psfmatching} and
\textsection\ref{ch:errors}).  We apply two corrections to the AUTO
flux to obtain better estimates of galaxies' total fluxes; these are
described below.  We will then quantify the effect and importance of
these corrections in the following section.

Even for a point source, any aperture that is comparable in size to
the PSF will miss a non-negligible amount of flux
\citep[e.g.][]{BertinArnouts, FasanoEtAl, K20survey, LabbeEtAl, BrownEtAl}.
\citet{BrownEtAl} have shown that fraction of light missed by the AUTO
aperture correlates strongly with total magnitude; this is simply due
to the fact that the AUTO aperture size correlates strongly with total
brightness.  \citet{LabbeEtAl} find that up to 0.7 mag can be missed
for some objects, and \citet{BrownEtAl} suggest that the systematic
effect at the faint end is $\sim 0.2$ mag.

It is therefore both appropriate and important to apply a correction
for missed flux laying outside the `total' aperture.  Following
\citet{LabbeEtAl}, we do this treating every object as if it were a
point source: using the empirical models of the PSF constructed as per
\textsection\ref{ch:psfmatching}, we determine the fraction of light
that falls outside each aperture as a function of its size and
ellipticity, and scale SExtractor's FLUX\_AUTO measurement
accordingly.  Since no object can have a growth curve which is steeper
than a point source, this is a minimal correction: it leads to a lower
limit on the total flux.

Further, we find that SExtractor's background estimation algorithms
systematically overestimate the background level, which also produces
a bias towards lower fluxes.  Because SExtractor does not allow the
user to turn off background subtraction when doing photometry (\cf\
detection), we are forced to undo SExtractor's background subtraction
for the final catalog, using the output BACKGROUND values, and the
area of the AUTO aperture.  We have done this only for the total $K$
fluxes; since we have performed our own background subtraction for the
NIR images (as described in \textsection\ref{ch:backsub}), undoing
SExtractor's background subtraction is equivalent to trusting our own
determination.  Note that, for the SED fluxes, we still rely on
SExtractor's LOCAL background subtraction algorithm, with PHOTO\_THICK
set to 48.

\subsection{Total Fluxes --- Validation} \label{ch:magsvalid}


\begin{figure*} \centering
\includegraphics[width=16cm]{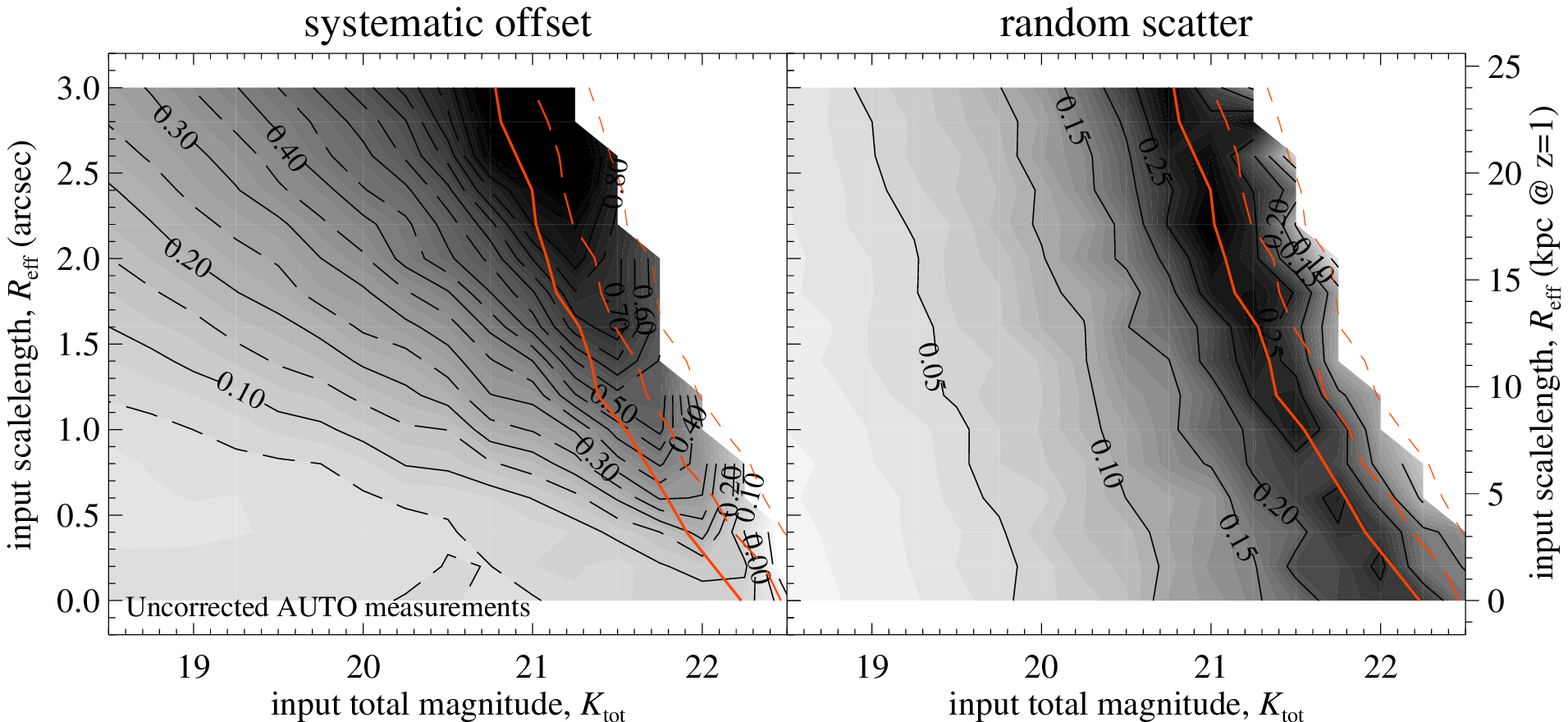}
\includegraphics[width=16cm]{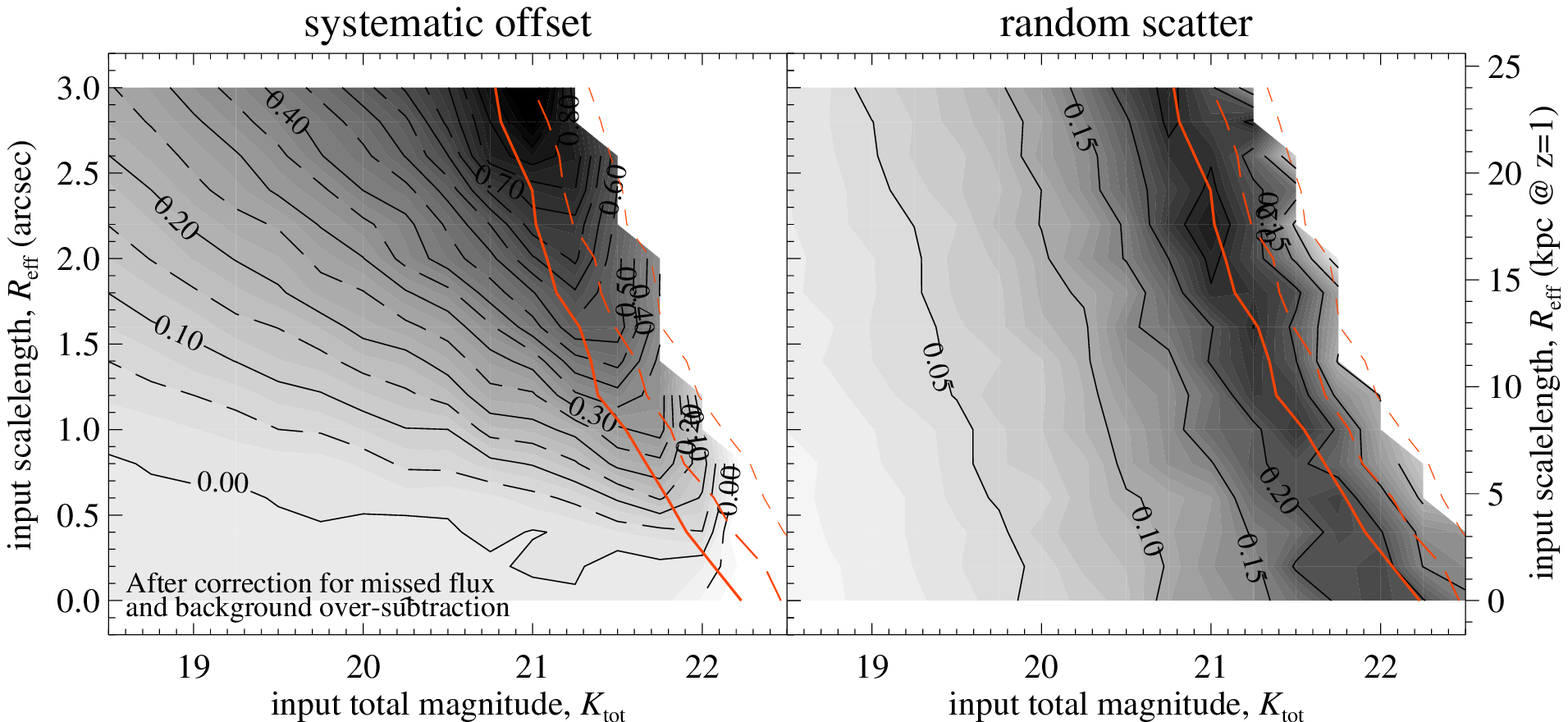}
\caption{Validating our total flux measurements---The systematic ({\em
    left panels}) and random ({\em right panels}) errors in the
    recovered fluxes of synthetic $R^{1/4}$--law sources introduced
    into the $1\farcs0$ FWHM $K$ science image, based on SExtractor's
    AUTO aperture, before ({\em upper panels}) and after ({\em lower
    panels}) applying corrections for missed flux and background
    oversubtraction.  The red lines in each panel show the approximate
    90 \% ({\em solid}), 50 \% ({\em long dashed}), and 10 \% ({\em
    short dashed}) completeness limits for $R^{1/4}$ law sources, as
    in Figure \ref{fig:completeness}.  As in Figure
    \ref{fig:completeness}, the results shown in this Figure are
    presented as upper limits on the systematic errors; both the
    random and systematic errors are significantly less assuming
    exponential profiles.  In order to account for flux laying beyond
    the AUTO aperture, we correct the flux of each source as if it
    were a point source; this is thus a minimal correction.  This
    correction reduces the systematic error in total fluxes by
    $\gtrsim 0.1$ mag for an $R^{1/4}$ profile, and from $\lesssim
    0.10$ mag to $\lesssim 0.01$ mag for point sources.  For the
    faintest sources, this correction also reduces the random error by
    as much as $0.05$ mag.  We also find that SExtractor's LOCAL
    background subtraction algorithm tends to over-estimate and
    over-subtract the background, leading to a systematic offset at
    the level of 0.03 mag.  With both of these corrections, the
    systematic errors in total fluxes for point sources are reduced to
    $\lesssim 2$ \% \label{fig:totalmags}}
\end{figure*}

Our overarching concern here is the correspondence between our
measured fluxes and the true total fluxes of real sources.  We have
tested our total flux measurements by checking our ability to recover
the known fluxes of large numbers synthetic sources, inserted into the
$1\farcs0$ FWHM $K$ science image as in
\textsection\ref{ch:completeness}.  The results of these tests are
shown in Figure \ref{fig:totalmags}.  In this Figure, we compare the
performance of SExtractor's AUTO measurement before ({\em upper
panels}) and after ({\em lower panels}) our corrections for missed
flux and background over-subtraction are applied.  In each case, the
contours show the systematic ({\em left panels}) and random ({\em
right panels}) errors in the recovered magnitude.  The red lines show
the approximate 90 \%, 50 \%, and 10 \% completeness limits for
$R^{1/4}$--law sources, as derived in
\textsection\ref{ch:completeness}. 

Further, in order to gauge the way these measurements are affected by
noise, we have performed several variations of this test.  In each
test we add the synthetic sources either to a noiseless image, or to
the actual $1\farcs0$ FWHM $K$ mosaic; we have trialled the four
possible permutations of using the noiseless or real image for
detection or measurement.  We briefly summarize the results of these
tests below.  

The reader wishing to avoid such a technical discussion of
SExtractor's photometry algorithms may wish to skip to
\textsection\ref{ch:seds} after noting that, comparing the upper and
lower panels of Figure \ref{fig:totalmags}, the effect of our two
corrections to the AUTO measurement is to reduce the systematic
underestimate of total fluxes by $\gtrsim 0.10$ mag.  For point
sources, the total flux is recovered to within 0.02 mag for $K <
22$. \looseness-1

\subsubsection{Missed Flux and Aperture Size Effects} \label{ch:auto}

In order to determine the bias inherent in the AUTO algorithm, we have
checked our ability to recover the fluxes of synthetic sources placed
in a noiseless image, using this image for both detection and
measurement.  For point sources, the photometric bias inherent in the
AUTO algorithm is $\lesssim 0.05$ mag for $K < 20.5$, but rises to
0.10 mag for $K = 22$.  It is also a strong function of $\reff$: at $K
= 21.5$, the AUTO aperture misses 0.12 mag for $\reff = 0\farcs4$, and
more than 0.25 mag for $\reff = 1\farcs0$.  Applying our `point
source' correction for missed flux reduces this bias to $< 0.02$ mag
for all $K < 22$ point sources; and, at $K = 21.5$, to 0.08 and 0.21
mag for $\reff = 0\farcs4$ and $1\farcs0$, respectively.

The above numbers indicate the bias inherent in the AUTO algorithm,
even for infinite signal--to--noise; considering synthetic sources
introduced into the real $K$ science image, we find that noise
exacerbates the problem.  For point sources, the mean offset between
the uncorrected AUTO and total fluxes are $\lesssim 0.05$ mag for $K <
20.0$, 0.10 mag for $K=21.5$ and 0.17 mag for $K = 22.0$.  For $K =
21.5$, the systematic offset is 0.16 mag for $\reff = 0\farcs4$ and
0.50 mag for $\reff = 1\farcs0$.  For $K = 22$, the average `point
source' correction for missed flux goes from 0.05 mag for true point
sources up to 0.10 mag for $\reff = 1\farcs0$, and 0.15 mag for $\reff
= 1\farcs5$.  After applying our correction for missed flux, the
photometric offset is reduced to $< 0.03$ mag for all point sources;
at $K = 21.5$, the numbers for $\reff = 0\farcs4$ and $\reff =
1\farcs0$ become 0.10 mag, and 0.35 mag, respectively.

As an aside, we have also looked at how noise {\em in the detection
image} affects the AUTO measurement, by using the real image (with
synthetic sources added), for detection, and using a noiseless image
for measurement.  The effect of noise in the detection image is to
induce scatter in the isophotal area, and so the AUTO aperture size,
at a fixed $\reff$ and $K$.  Applying a correction for missed flux
thus reduces the random scatter in the recovered fluxes of low surface
brightness sources, by eliminating the first order effects due to
aperture size; the random scatter in recovered fluxes is reduced by
$\sim 0.05$ mag for all $K \lesssim 21$ sources.  This can be seen in
Figure \ref{fig:totalmags}.

Also, as in \textsection\ref{ch:completeness}, note that the numbers
given above all apply to galaxies with an $R^{1/4}$ profile, and so
should be treated as approximate upper limits on the random and
systematic errors. We have performed the same test assuming
exponential profiles: the systematic error in the recovered flux is
less than 0.03 mag for all $K < 22$ and $\reff < 0\farcs6$.

\subsubsection{Background Oversubtraction}

Even after correcting for missed flux, and even for point sources,
SExtractor's photometry systematically underestimates the total fluxes
of synthetic sources.  At least part of this lingering offset is a
product of the LOCAL background subtraction algorithm.  This algorithm
uses a `rectangular annulus' with a user-specified thickness,
surrounding the quasi-isophotal detection region.  Any flux from the
source lying beyond this `aperture' (which may well be smaller than
the AUTO aperture) will therefore bias the background estimate
upwards, leading to oversubtraction, and so a systematic underestimate
of the total flux.

If we undo SExtractor's background subtraction\footnote{Again, note
that SExtractor does not allow the user to turn off background
subtraction for photometry.  In practice, we have undone SExtractor's
background subtraction using the output BACKGROUND value, multiplied
by the area of the AUTO aperture.  The AUTO aperture area is given by
KRON\_RADIUS$^2$ $\times~\pi~\times$ A\_IMAGE $\times$ B\_IMAGE.
Note, too, that we apply this correction {\em before} the missed flux
correction discussed in \textsection\ref{ch:auto}.}, then the
photometric offset for point sources is reduced to $< 0.02$ mag for
all $K < 22$.  The size of this correction is only weakly dependent on
source size and flux, varying from $\gtrsim -0.025$ mag for ($K$,
$\reff$) = (19, $0\farcs4$) to -0.038 for ($K$, $\reff$) = (22,
$0\farcs4$). 


\subsection{Multi-color SEDs} \label{ch:seds}

In order to maximize signal--to--noise for the faintest objects,
instead of measuring total fluxes in all bands, we construct
multi-color SEDs based on smaller, `color' apertures; we then use the
$K$ band total flux to normalize each SED.  

The `color' photometry is measured from $1\farcs5$ FWHM PSF-matched
images (see \textsection\ref{ch:psfmatching}), again using the
$1\farcs0$ FWHM $K$ mosaic as the detection image.  Specifically, we
use SExtractor's MAG\_ISO, again enforcing a minimum aperture size of
$2\farcs5$ diameter.  This limit is reached by essentially all objects
with $K > 21.5$, and essentially none with $K < 20.5$.  Note that,
even though the ISO aperture is defined from $1\farcs0$ FWHM $K$
mosaic, (after SExtractor's internal filtering; see
\textsection\ref{ch:detection}), all `color' measurements are indeed
made using matched apertures on $1\farcs5$ FWHM PSF-matched images.

In order to test our sensitivity to color gradients, we have verified
that $R_\mathrm{tot} = (R-K)_\mathrm{col}+K_\mathrm{tot}$, where
$R_\mathrm{tot}$ comes from using the $R$ band image in place of the
$K$ band image for detection and total flux measurement.  Particularly
for the brightest and biggest (and so, presumably, the nearest)
galaxies, the use of the ISO aperture is crucial in ensuring that this
is indeed true.

\subsection{Photometric Errors} \label{ch:errors}


\begin{figure}[b] 
\includegraphics[width=8.8cm]{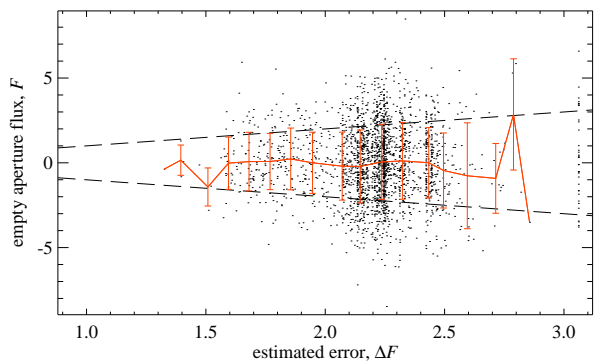}
\caption{Validating our error estimates---For the $JK$ imaging, we use
  the RMS maps output by xdimsum to account for spatial variations in
  the background noise level.  Although these maps are not accurate in
  an absolute sense, they do adequately map the relative variations in
  the noise; accordingly, we have normalized these RMS maps using the
  scatter in $2\farcs5$ diameter apertures placed on empty regions of
  the science image.  The points in this Figure show the integrated
  flux in each of 3000 of these apertures, $F$, as a function of the
  estimated error, $\Delta F$, derived using these RMS maps.  The red
  error bars show the RMS in $F$, binned by $\Delta F$.  The observed
  RMS in empty apertures agrees extremely well with the estimated
  errors based on the normalized xdimsum RMS map. \label{fig:errors}}
\end{figure}

Following, for example, \citet{LabbeEtAl}, \citet{GawiserEtAl}, and
\citet{QuadriEtAl}, we empirically determine the photometric
measurement uncertainties by placing large numbers of apertures on
empty or blank regions in our measurement images.  The principal
advantage of this approach is that it correctly accounts for
pixel--pixel correlations introduced in various stages of the data
reduction process (including interpolation during astrometric
correction and convolution during PSF matching).

For the `color' apertures, we have placed $2\farcs5$---$8''$
independent (\ie , non-overlapping) apertures at $10^4$ `empty'
locations, based on the combined optical ($BVR$) and NIR ($K$)
segmentation maps.  With this information, we can build curves of
$\sigma(A)$ for each band, where $\sigma$ is the measurement
uncertainty in an aperture with area $A$.  Similarly, for the `total'
apertures, which are somewhat larger, we have placed
$2\farcs5$--$12''$ independent apertures at 3500 `empty' locations on
the $1\farcs0$ FWHM $K$ detection mosaic, using only the NIR
segmentation map to define `empty'.  Note that since the `empty
aperture' photometry is done using SExtractor, in the same manner as
for our final photometry, the errors so derived also account for
random uncertainties due to, for example, errors in background
estimation, etc.

There is one additional layer of complexity for the ISPI bands: in
order to track the spatial variations in the `background' RMS, both
within and between subfields, we use the RMS maps produced during
mosaicking by xdimsum (see \textsection\ref{ch:xdimsum}).  While
these maps are not accurate in an absolute sense, they do adequately
map the shape of RMS variations across each subfield.  We have
therefore normalized these maps by the RMS flux in empty $2\farcs5$
apertures, and then combined them to construct a (re)normalized `RMS
map' for the full $30' \times 30'$ field.  Then, in practice, the
photometric uncertainty for a given object is estimated by taking the
median pixel value within the SExtractor segmentation region
associated with that object, corrected up from $2\farcs5$ to the
appropriate aperture size using the $\sigma(A)$ curves described
above.  

In Figure \ref{fig:errors}, we validate these error estimates by
showing the `empty aperture' fluxes, $F$, as measured in $2\farcs5$
diameter apertures, as a function of the photometric error, $\Delta
F$, estimated as above.  The line with error bars shows the mean and
variance of the `empty aperture' fluxes in bins of $\Delta F$; in
other words, the error bars show the {\em actual} error, plotted as a
function of the estimated error.  The agreement between the
photometric errors estimated using the RMS map, and the variance in
`empty aperture' fluxes is excellent.  This is more than just a
consistency check: while the RMS maps have been normalised to match
the variance in empty aperture fluxes on average, the fact that the
observed scatter scales so well with the predicted error demonstrates
that the RMS map does a good job of reproducing the spatial variations
in the noise.

For a Gaussian profile (\ie , a point source), and in the case of
uncorrelated noise, an aperture with a diameter 1.35 times the FWHM
gives the optimal S:N \citep{GawiserEtAl}.  Based on the `empty
aperture' analysis described in \textsection\ref{ch:errors}, the
$2\farcs5$ aperture size is slightly larger than optimal for a point
source in the $J$ ($1\farcs5$ FWHM) image.  For the $1\farcs0$ FWHM
$K$ detection image, the optimal aperture diameter for a point source
is $1\farcs33$; the S:N in a $2\farcs5$ diameter aperture is 25 \%
lower.  Using slightly larger apertures presumably increases S:N for
slightly extended sources, as well as reducing sensitivity to
systematic effects due to various classes of aperture effects (\eg ,
imperfect astrometric and PSF matching, etc.).

Within a $2\farcs5$ diameter aperture, the formal $5\sigma$ limits in
the $K$ band are 22.25 mag at an effective weight of 0.75, and 22.50
mag at an effective weight of 1.0.  Averaged across the image, the
$5\sigma$ limit is 22.42 mag; the limits for all bands are given in
Table \ref{tab:bands}.  For a point source, these limits can be
translated to total fluxes by simply subtracting 0.45 mag.


\section{Additional Checks on the MUSYC Calibration} \label{ch:extras}


\subsection{Checks on the Astrometric Calibration} \label{ch:atromcomps}

\begin{figure} \centering
\includegraphics[width=8.6cm]{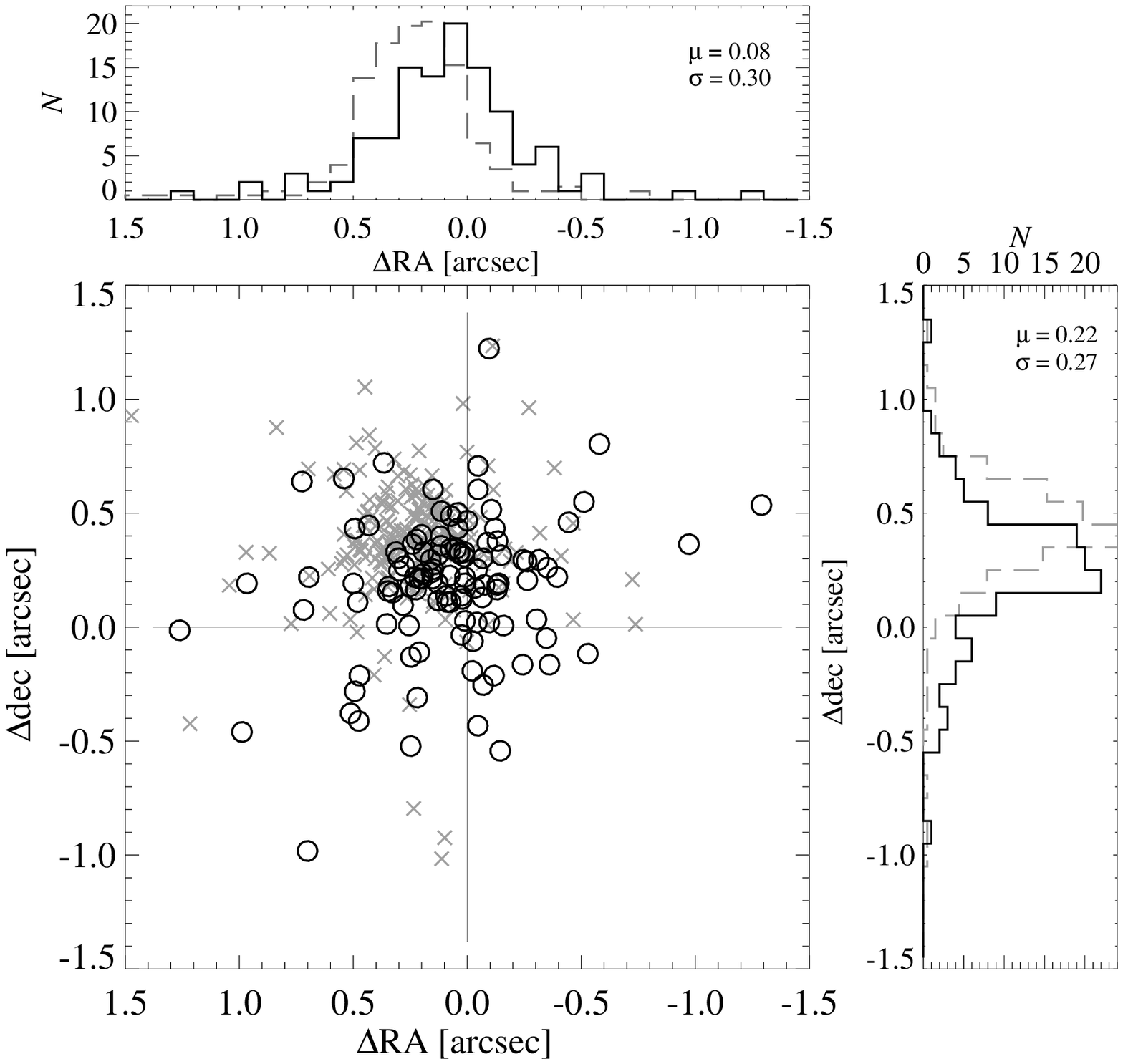}
\includegraphics[width=8.6cm]{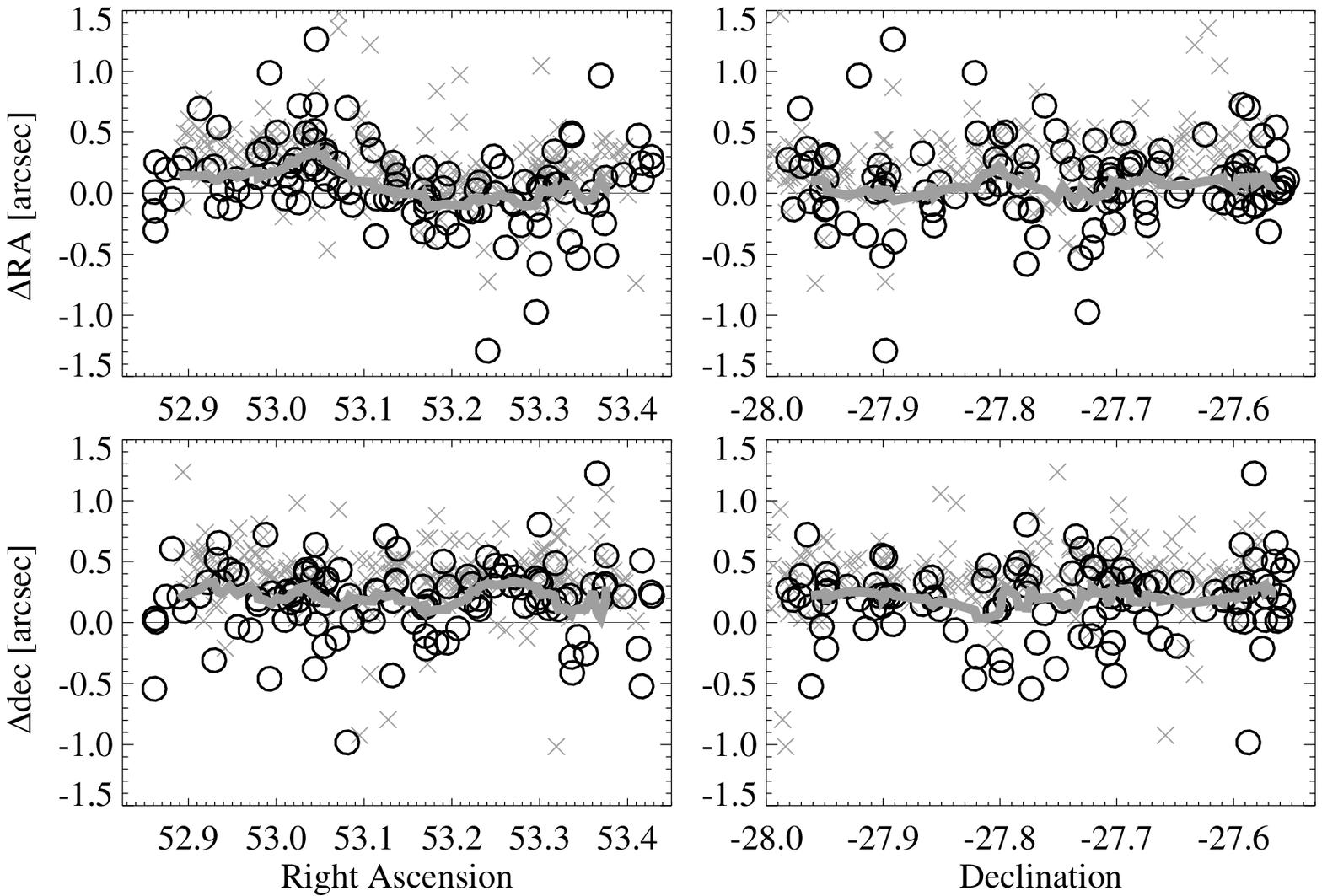}
\caption{Validating the MUSYC ECDFS astrometric calibration--- In the
  upper part of this Figure, we show the field-averaged astrometric
  comparison between the MUSYC ECDFS catalog, and the Yale/San Juan
  Southern Proper Motion (SPM) catalog v3.3 ({\em circles}), as well
  2MASS ({\em crosses}).  In the flanking panels, the solid (dashed)
  histograms show the distribution of RA/dec offsets with respect to
  the SPM (2MASS) catalogs; we also give the mean and RMS offset
  between MUSYC and SPM catalog positions.  In the lower part of this
  Figure, we show astrometric offsets as a function of position; in
  these panels, the solid grey lines show the median-filtered relation
  derived from the SPM points.  In comparison to the SPM catalog, the
  MUSYC astrometry is offset by $0\farcs23$ (0.87 pix); there is also
  evidence of an astrometric shear of $\lesssim 0\farcs3$ (1.1 pix) in
  the RA direction across the full field. \label{fig:astrom2}}
\end{figure}

In order to test the astrometric calibration of the MUSYC ECDFS
imaging, we have compared the cataloged position of sources from the
$K$--selected catalog with those from version 3.3 of the Yale/San Juan
Southern Proper Motion (SPM) catalog \citep{GirardEtAl2004}.  This
catalog is based on observations made using the 51 cm double
astrograph of Cesco Observatory in El Leoncito, Argentina.  For $V <
17$, the positional accuracy of the catalog is
$0\farcs04$---$0\farcs06$.

In Figure \ref{fig:astrom2}, we show an astrometric comparison for 113
objects common to the SPM and MUSYC catalogs; these objects are
plotted as black circles.  For this comparison we have selected
objects with $14 < V < 18$ and proper motions of less than 20 mas /
year.  All these objects have $14 < K < 18$; the median has $K = 16$
mag.

The systematic offset between SPM-- and MUSYC--measured positions,
averaged across the entire field, is $0\farcs079$ in Right Ascension
and $0\farcs222$ in declination; that is, a mean offset of
$0\farcs235$ (0.88 pix), 20$^\circ$ East of North.  For these sources,
the random error in the MUSYC positions is $0\farcs30$ and $0\farcs27$
in $x$ and $y$, respectively.

We have performed the same comparison for the 2MASS sources that were
used in the photometric calibration of the $K$ images; these objects
are shown in Figure 2 as the grey crosses.  The median $K$ magnitude
of these objects is 14.75 mag, considerably brighter than the SPM sources
used above.  In comparison to the 2MASS catalog, which has astrometric
accuracy of $\lesssim 0\farcs1$ for $K < 14$, we find a slightly larger
random offset: ($0\farcs22$, $0\farcs39$) in (RA, dec).  For these
sources, the random error in (RA, dec) is ($0\farcs22$, $0\farcs19$).

In the lower part of Figure \ref{fig:astrom2}, we plot the positional
offsets as a function of position accross the field.  In these panels,
the solid grey line shows the median-filtered relation between SPM--
and MUSYC--measured positions.  There appears to be a slight
astrometric shear in the RA direction at the $\lesssim 0\farcs3$ level
from the East to the West edge of the $K$ mosaic.  Otherwise, however,
the offsets are consistent with the direct shift of $0\farcs235$
derived above.


\subsection{Checks on the Photometric Calibration} \label{ch:photcomps}

\subsubsection{Comparison with FIREWORKS} \label{ch:goods}

\begin{figure} \centering
\includegraphics[width=8.6cm]{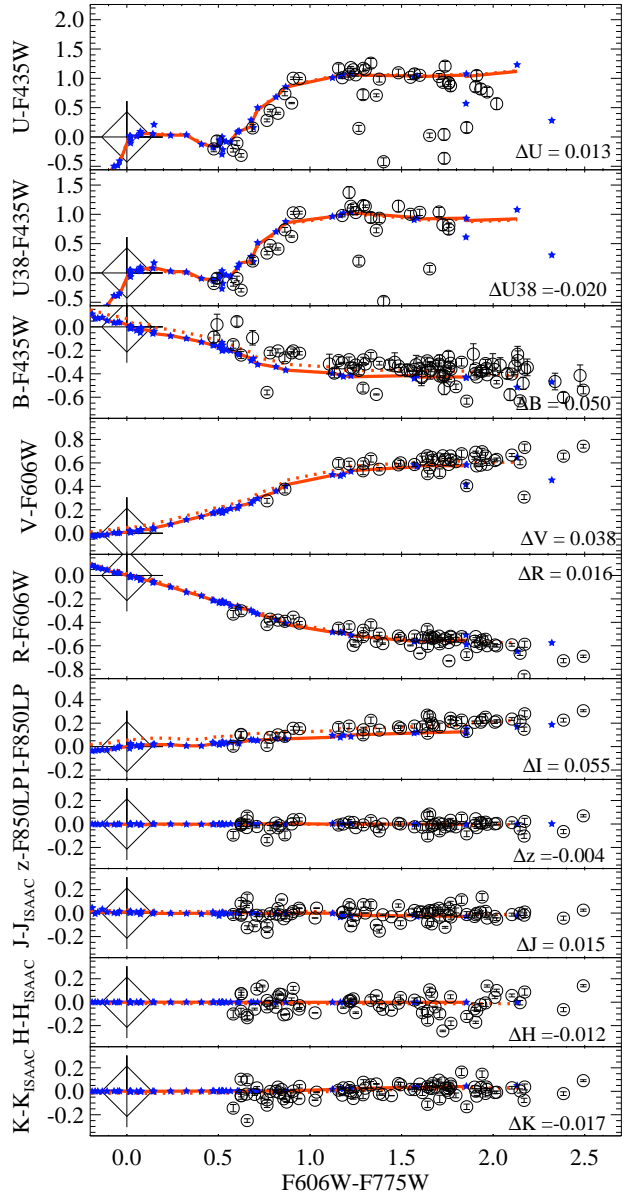}
\caption{Photometric comparison between GOODS and MUSYC in the CDFS,
  based on the FIREWORKS catalog \citep{WuytsEtAl} of the GOODS data
  ---In effect, these panels show the empirical color--transforms for
  stars between GOODS (ACS/ISAAC) and MUSYC (WFI/Mosaic-II/SofI/ISPI)
  filters, plotted as a function of ACS color. Note that for this
  comparison, we have adopted the Vega magnitude system, so that the
  stellar sequence must necessarily pass through the point (0, 0).  In
  each panel, the open circles with error bars represent the data; the
  errors shown here pertain only to the MUSYC photometry.  The closed
  blue stars show predicted stellar photometry based on the BPGS
  \citep{bpgs} stellar spectral atlas, convolved with the known filter
  curves; the solid red line shows a prediction for the stellar
  sequence in these color--color diagrams, obtained by median
  filtering the BPGS points.  For each filter, we derive a photometric
  offset by taking the mean difference, weighted by S:N in the MUSYC
  catalog, between the observed and predicted location of the stellar
  sequence in the $y$ direction.  These values are given in each panel
  (the `$\Delta$' having the sense of MUSYC--minus--GOODS); the dotted
  lines show the predicted stellar sequence offset by this amount.
  Particularly for the reddest bands, this comparison validates the
  MUSYC photometric calibration at the few percent level.
  \label{fig:goodsphotcomp} }
\end{figure}

In order to test our photometric calibration, we have compared our
catalog to the FIREWORKS catalog \citep{WuytsEtAl} of the GOODS-CDFS
region (the central $\sim 150 ~ \square''$ of our field), which
includes HST-ACS optical imaging, and significantly deeper NIR imaging
taken using ISAAC on the VLT.  Since the FIREWORKS catalog uses
different filters, we are forced to use stellar colors to make this
comparison.  The results of this comparison are shown in Figure
\ref{fig:goodsphotcomp}.  Each panel in this Figure shows the color--color
diagram for stars in terms of their FIREWORKS ($V_{606W}-I_{775W}$)
color, and a MUSYC--minus--FIREWORKS `color'.  In each panel, the
circles with error bars show the observations; these error bars apply
only to errors in the MUSYC photometry.

We have used spectra for luminosity class V stars from the BPGS
stellar spectral atlas \citep{bpgs} to generate predictions for where
the stellar sequence should lie in these diagrams.  These predictions
are the solid red lines in each panel; the small blue stars show the
predicted photometry for individual BPGS stars.  Note that, for the
purposes of this comparison, we have converted to the Vega magnitude
system, so that the stellar sequence necessarily passes through the
point (0, 0).

We calculate the photometric offset in each band as the S:N--weighted
mean difference between the observed stellar photometry and the
predicted stellar sequence.  These values are given in each panel; the
dashed red line is just the predicted stellar sequence offset by this
amount.  Our results do not change if we use the \citet{Pickles}
stellar atlas.  

Particularly for the NIR data, the absolute calibration of the MUSYC
and FIREWORKS data agree very well: typically to better than 0.03 mag.
In terms of the relative calibration across different bands, we see a
discrepancy between the $I$ and $z'$ band calibrations of $\Delta
(I-z')=0.05$ mag, as well as a discrepancy between the $U_{38}$ and
$B$ bands at the level of $\Delta(U_{38}-B) = -0.07$ mag.

\subsubsection{Comparison with COMBO-17} \label{ch:combo}


\begin{figure} \centering
\includegraphics[width=8.8cm]{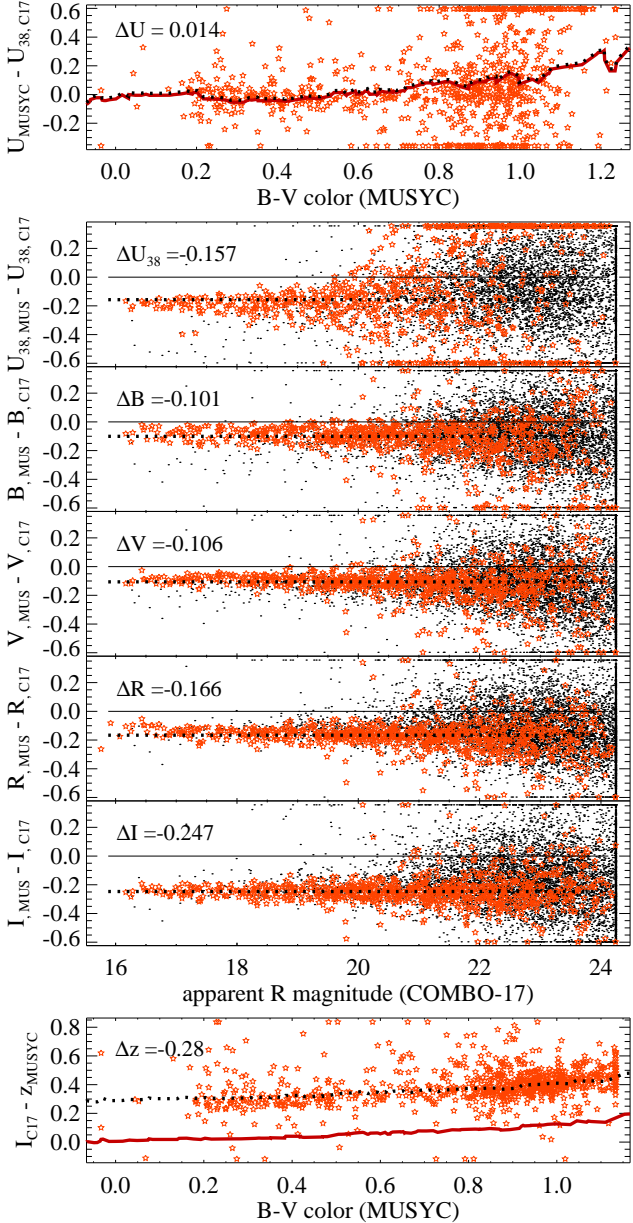}
\caption{Photometric comparison between COMBO-17 and MUSYC in the
  ECDFS --- Each panel shows the difference between the original
  COMBO-17 and MUSYC photometry; the red stars represent the observed
  photometry for stars, selected using COMBO-17's SED classification.
  In the central panels, which show direct object--by--object
  comparisons for those filters common to both MUSYC and COMBO-17, the
  black points show the same information for galaxies, plotted as a
  function of apparent $R$ magnitude in the COMBO-17 catalog.  The top
  and bottom panels compare the MUSYC $U$ and $z'$ bands to the
  COMBO-17 $U_{38}$ and $I$ bands, as a function of the MUSYC ($B-V$)
  color (in the Vega system), and based on synthetic photometry for
  main sequence stars.  The median MUSYC--minus--COMBO-17 photometric
  offset for each band is given in each panel; the dotted lines in
  each panel show the expected location of the stellar locus offset by
  this amount.  Note that while the COMBO-17 raw data is a subset of
  the MUSYC raw data, the data reduction and analysis pipelines are
  completely independent.  There are significant differences between
  the COMBO-17 and MUSYC photometry, due at least in part to
  photometric calibration errors in the original COMBO-17 catalog
  \citep{calibnote}.  Even after recalibrating the COMBO-17 photometry
  following \citep{calibnote}, however, significant differences
  remain: for $U_{38}BVRI$, the offsets are --0.014, --0.141, --0.109,
  --0.112, and --0.124 mag, respectively.
  \label{fig:combophot} }
\end{figure}

Although the COMBO-17 broadband $U_{38}BRVI$ imaging is a subset of
the raw data used to produce the MUSYC imaging, the data reduction and
analysis strategies used by each team are very different.  For
example, rather than a single measurement from a coadded image, the
COMBO-17 flux measurements are based on the coadding of many distinct
measurements from the individual exposures, and SED or `color'
measurements are made using adaptive, weighted `apertures', rather
than traditional (top-hat) apertures.  Direct, object--by--object
comparison between the two catalogs thus offers the chance to test
both the photometric calibration, and the methods used for obtaining
photometry.

The results of this comparison are shown in middle panels of Figure
\ref{fig:combophot}; these panels show the difference in the MUSYC and
COMBO-17 cataloged $U_{38}BVRI$ fluxes, plotted as a function of total
$R$ magnitude in the COMBO-17 catalog, $R_\mathrm{C17}$.  The
comparison is between total fluxes: \ie , $I_\mathrm{C17} =
R_\mathrm{tot, C17} + (I - R)_\mathrm{C17}$; $I_\mathrm{MUS} =
K_\mathrm{tot, MUS} + (I - K)_\mathrm{MUS}$.  We have also transformed
our data to the Vega magnitude system.  For the purposes of this
comparison, we distinguish between stars ({\em red stars}) and
galaxies ({\em black points}), on the basis of the COMBO-17 SED
classification; the results do not change significantly using $Bz'K$
selected stars or GEMS point sources.  We have used those stars with
$R < 21$ to identify differences in the two surveys' calibrations;
these offsets are given in each panel, and shown as the dotted black
lines.

There are significant differences between the MUSYC and original
COMBO-17 calibrations.  These are due to calibration errors in the
COMBO-17 catalog \citep{calibnote}. The original COMBO-17 calibration
was based on spectrophotometric observations of two stars, each of
which suggested different calibrations; in the end, the wrong star was
chosen.\footnote{Note that these calibration issues affect only the
ECDFS, and not the other three COMBO-17 fields, where multiple
calibration stars give consistent results \citep{calibnote}.} Partially
motivated by the comparison in Figure \ref{fig:combophot},
\citet{calibnote} have since revised the basic calibration of the
COMBO-17 ECDFS data using the other spectrophometric star, shifting
the $U_{38}BVRI$ calibration by --0.143, +0.040, +0.003, --0.054, and
--0.123 mag, respectively.  

We note that these rather large calibration errors do not have a huge
effect on the COMBO-17 redshift determinations \citep[][Paper
II]{calibnote}.  This is because the medium bands, which are key to
measuring break strengths and so choosing the redshift, are calibrated
with respect to the nearest broad band.  However, we show in Paper II
that the effect on derived quantities like restframe colors and
stellar masses is large.

After recalibration using the other spectrophotometric standard, the
MUSYC and COMBO-17 stellar colors agree at the level of a few
hundredths of a magnitude for $BVRI$; for $U_{38}$ a discrepancy
remains at the 0.1 mag level.  Moreover, a discrepancy in the overall
calibration remains, such that stars are 0.1 mag brighter in the MUSYC
catalog.  Our correction for missed flux accounts for 0.03 mag of this
offset; the source of the remaining 0.07 mag offset has not been
identified.

Secondly, notice that there are apparently different offsets for
galaxies and stars: even after matching the two surveys' calibrations
{\em for stars} using Figure \ref{fig:combophot}, {\em galaxies are
still fainter and bluer} in the COMBO-17 catalog than they are in
ours.  Quantitatively, the $U_{38}BVRI$ galaxy-minus-star offsets are
0.102, 0.020, 0.010, 0.067, and 0.088 mag, respectively.  Further,
excepting the $U_{38}$ band, the random scatter between the COMBO-17
and MUSYC galaxy photometry is 2---3 times greater than that for
stars.  It is difficult to say what might produce this effect, but the
effect persists even when we use our $R$ band image for detection and
measurement; that is, this is not a product of our measuring total
fluxes in $K$ rather than $R$.  We do not believe that the combination
of COMBO-17's smaller effective apertures and galaxy color gradients
can fully account for these effects.  For $R \gtrsim 21$, the
effective diameter of the ISO aperture is almost always smaller than
$2\farcs5$; for these objects the MUSYC photometry effectively uses
fixed apertures.  While the agreement between star and galaxy colors
is noticeably better for $R \lesssim 21$ using fixed $2\farcs5$
apertures to construct SEDs, it does not have a significant effect for
$R \gtrsim 21$, where the problem is greatest.

While we cannot directly compare our $U$ and $z'$ photometry to
COMBO-17, it is still possible to use stellar colors to check these
bands, as we have done for the FIREWORKS catalog.  This is shown in
the top and bottom panels of Figure \ref{fig:combophot}.  For the $z'$
band, this analysis suggests a possible discrepancy between the MUSYC
$I$ and $z'$ band calibrations of $\Delta (I-z')_\mathrm{MUS} = 0.03$
mag.  For the $U$ band, however, it suggests a discrepancy of $\Delta
(U-U_{38})_\mathrm{MUS} \sim 0.15$ mag.  While we have been unable to
identify the cause of this offset, we note both that the shape of the
observed and predicted stellar sequences do not obviously agree as
well for the $U$ band as for the $z'$, and also that the results of
both \textsection\ref{ch:goods} and \textsection\ref{ch:starcols} do
not support the notion of an offset of this size.  We do not believe
that this indicates an inconsistency in the calibrations of the $U$
and $U_{38}$ bands.

\subsubsection{Refining the Photometric Cross-Calibration using Stellar SEDs} 
\label{ch:starcols}

\begin{table} \begin{center}
\caption{Checks On the Photometric Calibration}
\begin{tabular*}{0.4\textwidth}{@{\extracolsep{\fill}}c c c c}
\hline \hline 
Band & \multicolumn{3}{c}{Photometric Offset with respect to} \\

& FIREWORKS & COMBO-17 & Stellar SEDs \\

(1) & (2) & (3) & (4) \\
\hline
$U$   & $+0.013$  & $+0.02$  & $-0.004$  \\
$U_{38}$ & $-0.020$  & $-0.15$  & $-0.051$  \\
$B$   & $+0.050$  & $-0.09$  & $-0.017$  \\
$V$   & $+0.038$  & $-0.09$  & $-0.006$  \\
$R$   & $+0.016$  & $-0.15$  & $+0.017$  \\
$I$   & $+0.055$  & $-0.23$  & $+0.023$  \\
$z'$  & $-0.004$  & $-0.27$  & $-0.011$  \\
$J$   & $+0.015$  &  ---     & $+0.032$  \\
$H$   & $-0.012$  &  ---     & $-0.032$  \\
$K$   & $-0.017$  &  ---     &    ---   \\
\hline \hline \label{tab:offsets}
\end{tabular*} \end{center}
\tablecomments{This Table summarizes the results of
  \textsection\ref{ch:photcomps}.  For each band (Col.\ 1), we give:
  (Col \.2): the photometric offset between MUSYC and the FIREWORKS
  \citep{WuytsEtAl} catalogs of the GOODS ACS and ISAAC imaging data;
  (Col.\ 3): the photometric offset between the MUSYC and COMBO-17
  \citep{WolfEtAl} optical imaging data; (Col.\ 4): the residuals from
  fitting stellar SEDs from the MUSYC catalog using main sequence
  stellar spectra from the BPGS atlas.}
\end{table}

In the construction of SEDs covering a broad wavelength range, the
relative or cross-calibration across all bands is at least as
important as the absolute calibration of each individual band.  As a
trivial example, if the zeropoints of two adjacent bands are out by a
few percent, but in opposite senses, this can easily introduce
systematic offsets in color on the order of 0.1 mag; the worry is then
that these apparent `breaks' might seriously affect photometric
redshift determinations.  This is a particular concern in the case of
the MUSYC ECDFS dataset, which incorporates data from four different
instruments, each reduced and calibrated using quite different
strategies. \looseness-1

We have therefore taken steps to improve the photometric
cross-calibration of the MUSYC ECDFS data.  The essential idea here is
to take a set of objects whose SEDs are known {\em a priori} (at least
in a statistical sense) and to ensure agreement between the observed
and expected SEDs.  Stars are, in fact, ideal for this purpose, since
they form a narrow `stellar sequence' when plotted in color--color
space: at least in theory, and modulo the effects of, \eg , metalicity,
a star's (\cf\ a galaxy's) full SED can be predicted on the basis of a
single color.

Our method is as follows.  We begin with a set of more than 1000
objects with unambiguous `Star' classifications in the COMBO-17
catalog, of which nearly 600 have photometric S:N $\gtrsim 10$ in $K$,
and are unsaturated in all MUSYC bands.  Again, our results do not
change if we use $Bz'K$ selected stars or GEMS point sources.  Using
EAZY (see \textsection\ref{ch:eazy} for a description), we fit the
objects' photometry with luminosity class V stellar spectra from the
BPGS stellar spectral atlas as a template set, and the redshift fixed
to zero.  Note that, by default, EAZY includes a 0.05 mag systematic
error on each SED point, added in quadrature with the measurement
uncertainty.

Using the output $\chi^2$ to discard objects whose SEDs are not
consistent with being a main sequence star, we can then interpret the
median residual between the observed and best-fit photometry as being
the product of calibration errors, and so refine the photometric
calibration of each band to ensure consistency across all bands.
Specifically, given the photometric errors, we use $\chi^2$
minimization to determine the zeropoint revision.  

The zeropoint revisions derived in this way are small; $\lesssim 0.05$
mag in all cases.  The exact revisions are given in Table
\ref{tab:offsets}.  Across the WFI data, there appears to be an offset
that is roughly monotonic between the $U_{38}$ and $R$ bands, where
the offset in $U_{38} - I$ is $-0.054$ mag; \cf\ 0.055 mag from
the comparison to the FIREWORKS catalog.  Similarly, there is an
apparent inconsistency between the $I$ and $z'$ calibrations, such
that the offset in $(I-z')$ is 0.03 mag; \cf\ 0.05 mag from the
comparison to FIREWORKS.

The crux of this method is that whatever zeropoint discrepancies exist
do not affect the choice of the best fit template in a systematic way.
For example, a large offset in the $U$ bands or a
wavelength--dependent offset might lead to stars being fit with
systematically bluer or redder template spectra, so biasing the
derived photometric offsets.  In this sense, it is reassuring that the
derived offsets are small, and comparable to the quoted uncertainties
on the photometric calibration.  Further, we note that we get very
similar results if we increase the systematic uncertainty used by EAZY
to 0.10 mag.


\begin{figure*} \centering
\includegraphics[width=17.8cm]{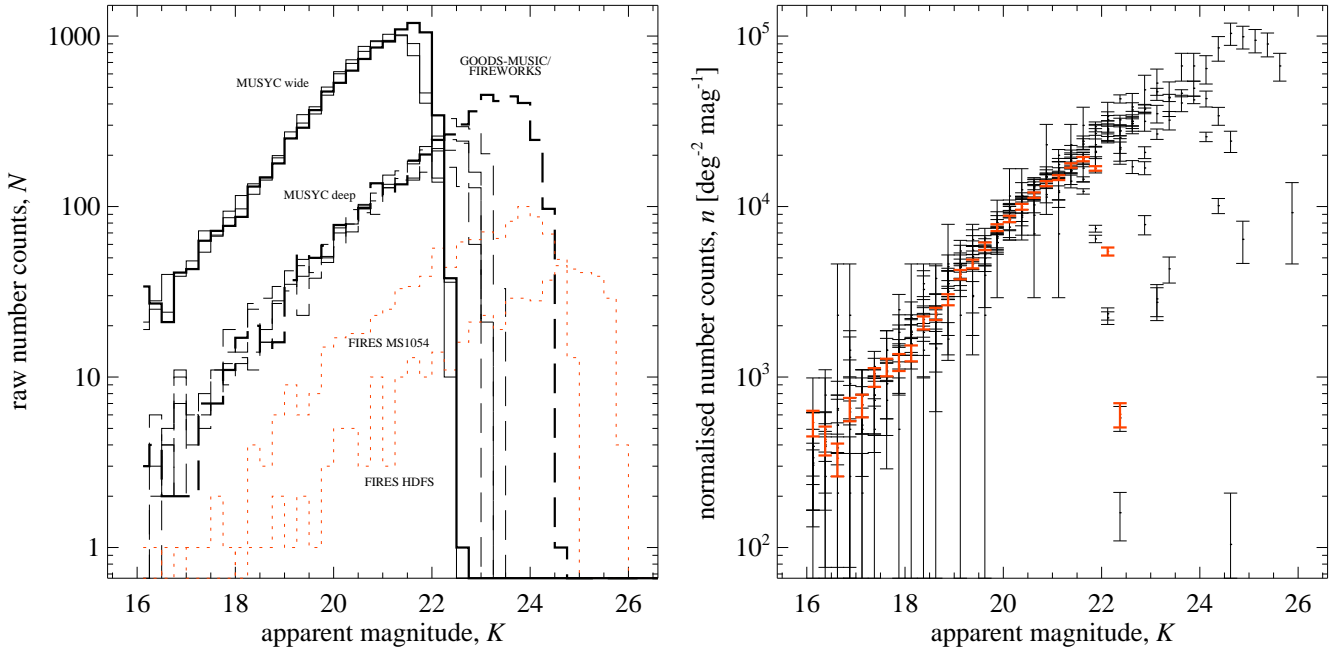}
\caption{$K$ band apparent magnitude number counts, comparing the
  MUSYC ECDFS catalog to other $K$--selected catalogs---{\em Left
  Panel:} the raw numbers of detected sources, in bins of $K$
  magnitude for the ECDFS ({\em heavy solid histogram}), in comparison
  to: the other MUSYC wide fields \citep[][{{\em light solid
  histograms}}]{BlancEtAl}; the MUSYC deep fields \citep[][{{\em light
  dashed histograms}}]{QuadriEtAl}; the FIREWORKS catalog of the
  GOODS-CDFS data \citep[][{{\em heavy dashed histogram}}]{WuytsEtAl};
  and the two FIRES fields \citep[][{{\em red dashed
  histograms}}]{LabbeEtAl, NMFS-FIRES}.  {\em Right panel:} The
  normalized number counts for the same collection of datasets; the
  MUSYC ECDFS data are highlighted ({\em heavy red points}).  At a
  fixed $K$ magnitude, while the GOODS region of the ECDFS has
  approximately 80 \% as many sources as are found in the ECDFS as a
  whole, in comparison to the other MUSYC wide fields, the ECDFS is
  underdense at the level of $\sim 5$ \%.
\label{fig:nocounts}} \end{figure*}

Given the agreement between the results of the external comparison to
FIREWORKS and those from the internal consistency check on stellar
colors, we have chosen to adopt the zeropoint revisions suggested by
this stellar colors exercise.  With these revisions, we believe that
our photometric calibration is accurate, in both an absolute and a
relative sense, to the level of a few hundredths of a magnitude.


\section{Number Counts} \label{ch:nocounts}

As a very basic comparison between our catalog and other $K$--selected
catalogs, Figure \ref{fig:nocounts} shows the number of detected
galaxies as a function of total apparent $K$ magnitude.  Note that all
the catalogs shown apply a similar correction for flux missed by
SExtractor's AUTO measurement.  The left panel of this figure shows
the raw number counts; the right shows the number counts normalized by
area.  In both panels, it can be seen that our number counts drop off
for $K \gtrsim 22$; our catalog is nearly, but not totally, complete
for $K = 22$. 

The overall agreement between these different catalogs is very good.
Assuming that the calibration of all catalogs is solid, and looking at
the left panel of Figure \ref{fig:nocounts}, it can be seen that the
ECDFS is slightly underdense --- at the level of 4---6 \% for $17.5 <
K < 21.5.$ --- in comparison to the two other MUSYC wide NIR selected
catalogs \citep{BlancEtAl}.  Conversely, the ECDFS number counts can
be matched to the other two wide catalogs by adjusting the ECDFS $K$
photometric calibration by $-0.06$ or $-0.09$ mag.

In comparison to the number counts from the FIREWORKS catalog of the
GOODS CDFS region, the GOODS region contains approximately 18 \% fewer
sources per unit area than the ECDFS as a whole.  Even after matching
the MUSYC ECDFS $K$ calibration to the FIREWORKS catalog (see
\textsection\ref{ch:goods}), the GOODS region remains underdense by
16 \% in comparison to the ECDFS. 


\section{Photometric Redshifts} \label{ch:derived}

\subsection{Star/Galaxy Separation} \label{ch:stargal}

\begin{figure*} \centering
\includegraphics[width=18cm]{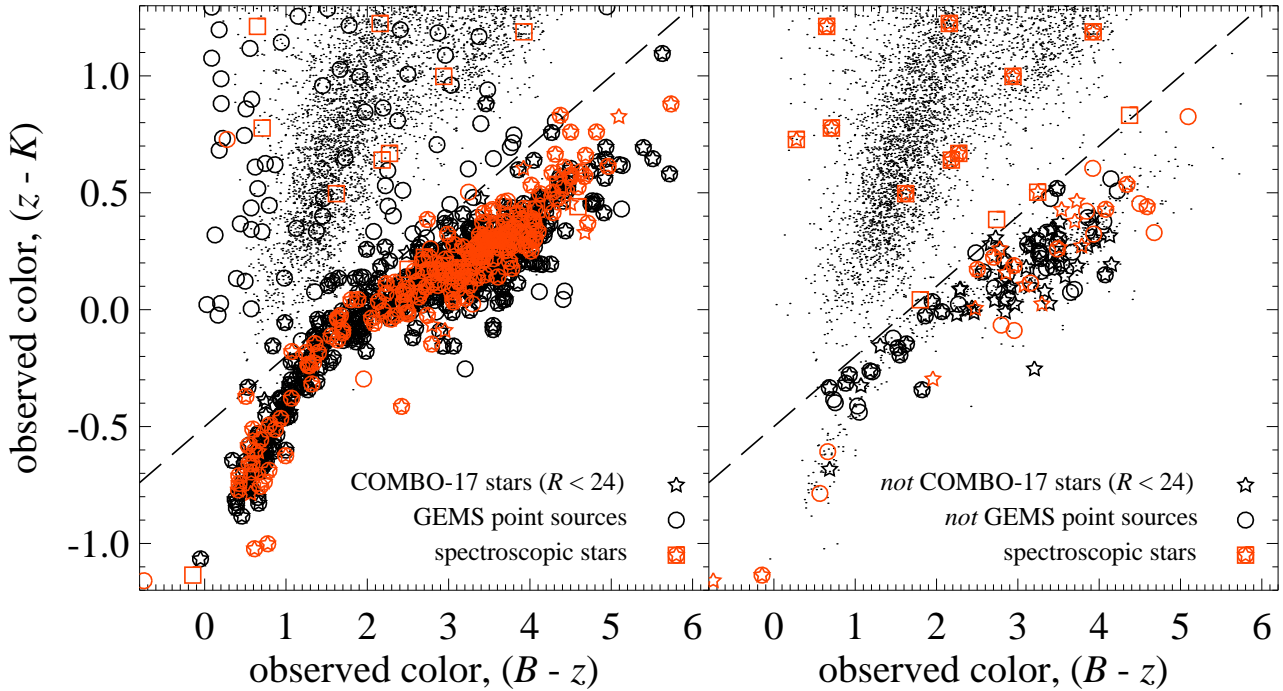}
\caption{Stellar identification using $Bz'K$ colors---In each panel,
  we show the $Bz'K$ diagram for sources in the MUSYC ECDFS catalog
  (black points), and compare our $Bz'K$ star selection ({\em dashed
  line}) to other complimentary stellar classifications: \viz\
  SED--classified `stars' from the COMBO-17 survey \citep[][{\em open
  stars}]{WolfEtAl}, GEMS point sources \citep[][{\em open
  circles}]{HausslerEtAl}, and spectrally classified stars ({\em open
  squares/red symbols}).  The left panel shows the agreement between
  $Bz'K$ selection and these other indicators; in the right panel we
  show where $Bz'K$--selection disagrees with other indicators.  So,
  for example, circles in the left panel show all GEMS point sources,
  whereas in the right panel they show those $Bz'K$--selected `stars'
  that are {\em not} GEMS point sources.  In the either panel, the
  stellar sequence in $Bz'K$ color space can be seen to be isolated by
  $\gtrsim 0.1$ mag in ($z'-K$) from deep field galaxies.  This
  includes QSOs, which can be seen in the left panel as GEMS point
  sources scattered throughout the galaxy population.  Although there
  are a handful of spectrally classified stars lying well outside the
  $Bz'K$ stellar selection region (open squares in the left panel),
  these objects are neither COMBO-17 `stars' nor GEMS point sources
  (stars and circles in the right panel); \ie\ the spectral
  classification is wrong.  Of the $Bz'K$--selected stars which are
  not GEMS point sources (circles in the right panel), roughly half
  are faint stars superposed over a diffuse background galaxy, and
  roughly half are faint galaxies whose photometry is significantly
  affected by a bright, nearby star.
  \label{fig:stars}}
\end{figure*}

We separate stars and galaxies from within the MUSYC ECDFS catalog on
the basis of their $Bz'K$ colors.  The $Bz'K$ diagram is known as a
means of selecting moderate redshift ($z \gtrsim 1.4$) galaxies
\citep{DaddiEtAl-BzK}, but can also be used as a efficient means of
distinguishing stars from galaxies \citep[see, \eg , ][]{GrazianEtAl,
BlancEtAl}.  In Figure \ref{fig:stars}, we evaluate the performance of
this criterion in comparison to the stellar SED classification from
COMBO-17 \citep{WolfEtAl}, as well as to a catalog of point sources
from GEMS \citep{HausslerEtAl}. \looseness-1

Both panels of Figure \ref{fig:stars} show the $Bz'K$ diagram for the
MUSYC ECDFS catalog ({\em black points}); the $Bz'K$ stellar
selection line:
\begin{equation}
(z'-K) \le 0.3 ~ (B-z') - 0.5 ~ ,
\end{equation}
is shown as the dashed line.  In total, from the main $K < 22$ sample,
755 sources are selected as stars on the basis of their $Bz'K$ colors.
The left-hand panel of Figure \ref{fig:stars} shows where $Bz'K$ star
selection agrees with other indicators; the right-hand panel shows
where there is disagreement.  For instance, on the left, the
star-shaped symbols show objects that are classified as `stars' by
COMBO-17; on the right, they represent those $Bz'K$--selected `stars'
which are {\em not} classified as such in the COMBO-17 catalog.
Similarly, the circles refer to point sources in the GEMS catalog.
In both panels, objects that have been spectrally identified as stars
are highlighted in red.  In either panel, the stellar sequence is
immediately obvious and, for a given ($B-z'$) color, can be seen to be
separated from the galaxy population by at least a few tenths of a
magnitude in $(z'-K)$.

Looking at the left panel, there is near complete overlap between
COMBO-17's star classification and $Bz'K$ selection: only a very few
COMBO-17 `stars' lie above the $Bz'K$ selection line.  There are a few
dozen GEMS point sources found above the $Bz'K$ selection line.  In
the MUSYC and GEMS optical images, some are clearly non--circular, and
only a few show diffraction spikes; these appear to be compact, un--
or barely--resolved galaxies.  Note, too, that this region of the
$Bz'K$ diagram is sparsely populated by X-ray sources \citep[\ie\
QSOs;][]{DaddiEtAl-BzK, GrazianEtAl}.

There are also a handful of objects that are spectroscopically
identified as stars, which also fall above the $Bz'K$ star selection
line.  With one exception, however, these objects are not GEMS point
sources (squares in the left panel; circles in the right); neither are
they classified as stars by COMBO-17 (squares in the left panel; stars
in the right).  These are, therefore, probably erroneous spectral
classifications.  There are no spectroscopic galaxies that lie in the
stellar region of the $Bz'K$ diagram.

Turning now to the right panel, there are 66 $Bz'K$--selected `stars'
which do not appear in the GEMS point source catalog.  A handful of
these simply did not receive GEMS coverage.  Of the rest, visual
inspection shows these sources to be, in roughly equal proportions,
faint stars superposed over a faint, background disk galaxy, or faint
galaxies whose photometry is significantly affected by a nearby bright
star.  There are also 76 $Bz'K$--selected `stars' which are not
classified as such in the COMBO-17 catalog.  In $(J-K)$--$K$
color--magnitude space, these objects almost all have $(J-K) < 0$ and
$K < 21$; this would suggest that these are faint stars misclassified
by COMBO-17.


\subsection{Photometric Redshifts --- Method} \label{ch:eazy}

\begin{table*} \centering 
\caption{Summary of the Contents of the Photometric Redshift Catalog}
\begin{tabular*}{0.95\textwidth}{@{\extracolsep{\fill}}l l p{12 cm}}
\hline\hline \label{tab:photzcat}
\\
Column No. & Column Title & Description \\
\\
\hline\hline
\\
1 & id & Object identifier, beginning from 1, as in the photometric catalog \\

2 & z\_spec & Spectroscopic redshift determination, where available, as given in the photometric catalog \\

3, 4 & z\_a, chi\_a & Maximum likelihood redshift, allowing
non-negative combinations of all six of the default EAZY templates,
and the $\chi^2$ value associated with each fit \\

5, 6 & z\_p, chi\_p & As above, but with the inclusion of a $K$
luminosity prior \\

7, 8 & z\_m1, z\_m2 & Probability--weighted mean redshift, without and
with the inclusion of a $K$ luminosity prior, respectively; we
recommend the use of the z\_m2 redshift estimator. \\

9---14 & l68, u68, etc. & Lower and upper limits on the redshift at 68, 95, and 99 \% confidence, as computed from the same posterior probability distribution used to calculate z\_m2 \\

15 & odds & The fraction of the total integrated probability within
$\pm 0.2$ of the z\_m2 value \\

16 & qz & The $Q_z$ figure of merit proposed by \citet{eazy},
calculated for the z\_m2 value \\

17 & nfilt & The number of photometric points used to calculate all of
the above \\

\\
\hline
\hline
\end{tabular*}
\end{table*}


The basic idea behind photometric redshift estimation is to use the
observed SED to determine the probability of an object's having a
particular spectral type, $t$ (drawn or constructed from a library of
template spectra), and being at a particular redshift, $z$: \ie\
$p(z,~t|\mathrm{SED})$.  We have derived photometric redshifts for
every object in the catalog using a new photometric redshift code
called EAZY \citep[Easy and Accurate $\zphot$s from Yale; for a more
detailed and complete discussion, see][] {eazy}.  EAZY combines many
features of other commonly used photometric redshift codes like a
Bayesian luminosity prior \citep[\eg\ BPZ;][]{bpz} and template
combination \citep{RudnickEtAl2001,RudnickEtAl2003} with a simple user
interface based on the popular hyperz code \citep{hyperz}.  Novel
features include the inclusion of a `template error function'; a
restframe wavelength dependent systematic error, which down-weights
those parts of the spectrum like the restframe UV, where galaxies show
significant scatter in color--color space.  Moreover, the user is
offered full control over whether and how these features are employed.


\begin{figure*} \centering
\includegraphics[width=18cm]{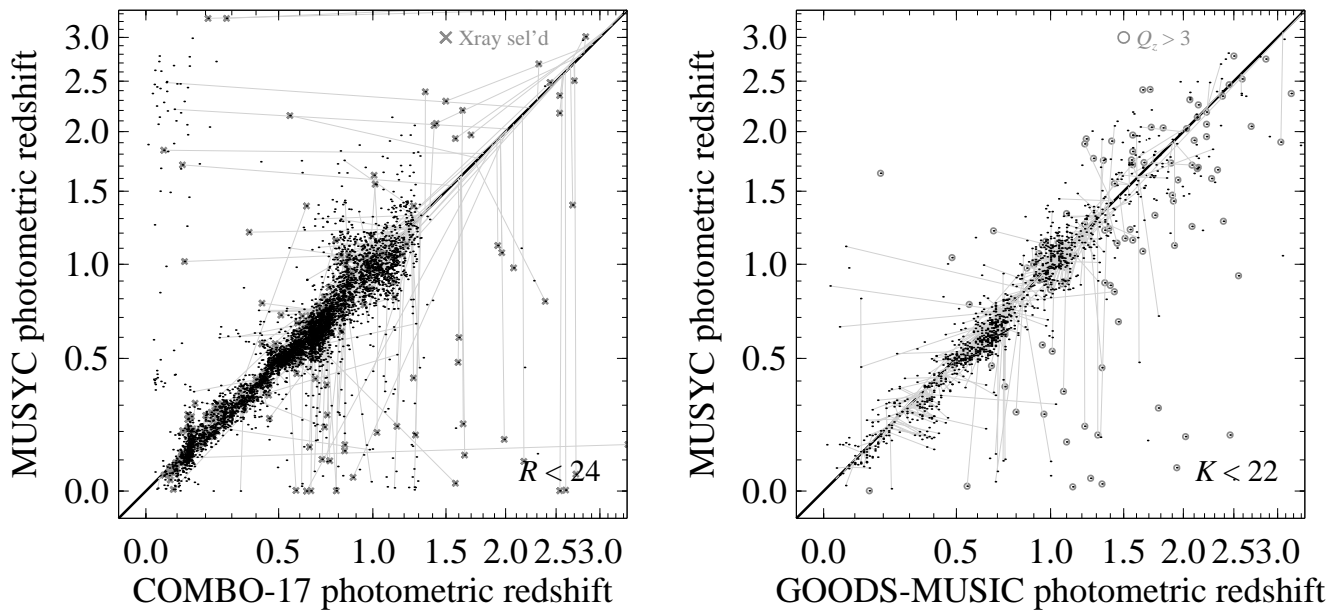}
\caption{Validating the MUSYC ECDFS photometric redshifts---Each panel
  shows an object--by--object comparison between the MUSYC photometric
  redshift, and that from COMBO-17 \citep[][{{\em left
  panel}}]{WolfEtAl}, and the GOODS-MUSIC catalog of the GOODS-CDFS
  data \citep[][{{\em right panel}}]{GrazianEtAl}.  In order to
  discriminate between the two $\zphot$s where there is disagreement,
  where a robust spectroscopic redshift determination is available
  (see Appendix \ref{ch:speczs}), the red lines connect each point in
  ($\zphot$, $\zphot$) point to the point ($\zspec$, $\zspec$);
  vertical lines thus indicate catastrophic errors in the MUSYC
  $\zphot$s, where horizontal lines show catastrophic failures in the
  COMBO-17/GOODS-MUSIC $\zphot$s.  COMBO-17 suffers from a few
  different classes of systematic effects, owing principally to the
  lack of NIR data.  Note, however, that very few spectroscopic
  redshifts are available for these objects---{\em these effects would
  not be noticeable in a} $\zspec$---$\zphot$ {\em diagram.}  In this
  panel, X-ray selected sources are marked with a cross; within this
  $R < 24$ sample, X-ray-selected sources are roughly three times as
  likely to have $|\Delta z|/(1+\zspec) > 0.1$ (see also Appendix
  \ref{ch:speczs}).  In comparison to GOODS-MUSIC, the MUSYC $\zphot$s
  have a slightly greater number of catastrophic outliers, such that
  the MUSYC $\zphot$ is far too low; again, many of these objects are
  X-ray sources.  In this panel, objects with poorly constrained
  photometric redshifts (\ie\ $Q_z > 3$) are marked with a circle;
  these objects are roughly twice as likely to have
  $|(z_\mathrm{MUS}-z_\mathrm{GDS})|/(1+z_\mathrm{GDS}) > 0.2$.  The
  overall agreement between the two redshift determinations is really
  very good, especially moving towards the `redshift desert' at
  $\zphot \gtrsim 1.5$.
\label{fig:photz} }
\end{figure*}

Another key difference is that objects are assigned redshifts by
taking a probability weighted integral over the full redshift grid
(\ie\ marginalizing over the posterior redshift probability
distribution), rather than, for example, choosing the single most
likely redshift.  (Although again the user is given the choice of
which estimator to use.)  EAZY also outputs 68/95/99 \% confidence
intervals, as derived from the typically asymmetric $p(z)$.  EAZY thus
outputs meaningful and reliable photometric redshift errors, including
the effects of `template mismatch'; \ie\ degeneracies between the
redshift solution and the spectral type.  By Monte Carlo'ing our
catalog (\ie\ reanalyzing many Monte Carlo realizations of our
photometry, perturbed according to the photometric errors), we have
verified that the EAZY $p(z)$ does in fact provide a good description
of the redshift uncertainties due to photometric errors. \looseness-1

We have adopted EAZY's default parameter set for our redshift
calculations.\footnote{In Paper II, we present a number of variations
on the photometric redshift computation described here; in relation to
Paper II, the redshifts described here correspond to the `default
analysis' in Paper II.}  That is, we use a library of six template
spectra, allowing non-negative linear combinations between these basis
templates, and including an apparent $K$ magnitude prior, $p(z|K)$,
and using the default EAZY template error function.

Both the base template set and the $K$ prior have been derived by
\citet{eazy} using synthetic photometry from the semi-analytic model
of \citet{DeLuciaBlaizot}, which is in turn based on the Millenium
Simulation \citep{Springel}.  The motivation for this approach is to
approximately account for the full diversity in $0 < z \lesssim 4$
galaxies' SEDs due to differences in their individual star formation
and assembly histories.  The $K$ prior is constructed directly from
the \citet{DeLuciaBlaizot} simulation.

In order to derive the base template set, \citet{eazy} have applied
the non-negative matrix factorization (NMF) algorithm of
\citet{blanton:photz}, to this synthetic catalog.  In essence, this
algorithm takes a large template library and distills from it a
reduced set of basis templates that best describe the full range of
`observed' photometry.  For this purpose, \citet{eazy} have used the
template library used by \citet{GrazianEtAl} to generate photometric
redshifts for the GOODS-MUSIC catalog.  This library consists of $\sim
3000$ P\'egase synthetic spectra with a variety of dust obscuration,
star formation histories, and ages.  In additional to the five base
templates output by the NMF algorithm, \citet{eazy} also include one
young, dusty template ($t$ = 50 Myr; $A_V = 2.75$), to compensate for
the lack of dusty galaxies in the \citet{DeLuciaBlaizot} similuation.

\citet{GrazianEtAl}, using their full template library, achieved a
photometric redshift accuracy of $\sigma_z = 0.045$ for their
GOODS-MUSIC catalog of the GOODS ACS-ISAAC-IRAC data.  For the same
data, and using the default setup described above, the EAZY
photometric accuracy is $\sigma_z = 0.036$.  This represents the
current state of the art for photometric redshift calculations based
on broadband photometry.

Table \ref{tab:photzcat} gives a summary of the information contained
within the photometric redshift catalog.  Note that when computing
photometric redshifts, we only use photometry with an effective weight
of 0.6 or greater.  In addition to the basic EAZY output, we have
included two additional pieces of information.  The first is simply a
binary flag indicating whether or not each object is classified as a
star on the basis of its $Bz'K$ colors.  The second is the figure of
merit proposed by \citet{eazy}:
\begin{equation}
Q_z(\zphot) = \frac{\chi^2}{N_\mathrm{filt}-3} ~
\frac{z^{99}_\mathrm{up} - z^{99}_\mathrm{lo}}{p_{\Delta z=0.2}} ~ .
\end{equation}
This quantity combines the $\chi^2$ of the fit at the nominal
redshift, the number of photometric points used in the fit,
$N_\mathrm{filt}$, the width of the 99 \% confidence interval,
$(z^{99}_\mathrm{up} - z^{99}_\mathrm{lo})$, and the fractional
probability that the redshift lies within $\pm 0.2$ of the nominal
value, ${p_{\Delta z=0.2}}$; all of these quantities are output by
EAZY by default.  \citet{eazy} have shown that a cut of $Q_z >
2$---3 can remove a large fraction of photometric redshift outliers.


\subsection{Photometric Redshifts --- Validation} \label{ch:photz-valid}

In Appendix \ref{ch:speczs}, we describe both the spectroscopic
redshift determinations that we have compiled for objects in the
ECDFS, and show the $\zphot$--$\zspec$ agreement for individual
$\zspec$ samples.  For all `secure' redshift determinations, the
random and systematic photometric redshift error is $\sigma_z$ = 0.036
and $\mathrm{med}[\Delta z/(1+z)] = -0.025$.  In comparison to
spectroscopic redshifts from the K20 survey, which is highly
spectrally complete in the magnitude regime in which we are operating,
the random error is $\sigma_z = 0.033$, with an outlier fraction of
less than 5 \%.  (Here, we define the outlier fraction as the relative
number of sources for which $\Delta z/(1+z) > 5 \sigma_z$.)  We also
draw particular attention to the excellent agreement between our
photometric redshifts and the spectroscopic determinations for the
sample of \citet{vdwel05}, which is a sample of 28 early type, red
sequence galaxies at $z \sim 1$; we find $\sigma_z = 0.022$, with no
outliers, and essentially no systematic offset.  For comparison, the
overall photometric redshift accuracy of the COMBO-17 survey for our
$\zspec$ comparison sample, but limited to $\zspec < 1$, is $\sigma_z
= 0.020$. 

However, we also show in Appendix \ref{ch:speczs} that none of the
available spectroscopic samples is particularly representative of the
MUSYC ECDFS sample.  In particular, in almost all cases there is a
correlation between redshift security and $(J-K$) color, such that
redshift determinations for blue galaxies tend to be more secure, and
so these galaxies are over-represented among MUSYC ECDFS galaxies.
Even the K20 sample, which is 92 \% complete for $K^\mathrm{(Vega)} <
20$, does not probe the reddest galaxies in our sample, presumably
because they are too rare to be found in that survey's rather small
area.  There is, therefore, the very real danger that looking only at
the $\zspec$---$\zphot$ agreement provides a false sense of security
\citep[see also][]{eazy}, since there are comparatively few
$\zspec$s available for the faintest and reddest galaxies in the
catalog---especially given that these are the main objects of
interest.

For this reason, we have compared our photometric redshifts to those
from COMBO-17 \citep{WolfEtAl} and GOODS-MUSIC \citep{GrazianEtAl};
the results of this comparison are shown in Figure \ref{fig:photz}.
While these comparisons are extremely useful for identifying
systematic differences between different $\zphot$ solutions, without
spectroscopic redshifts as a referent, they cannot be used to decide
which is `better' in the case of a disagreement.  To this end, the red
lines in each panel of this Figure show the spectroscopic redshifts
(where available) by connecting the ($\zphot$, $\zphot$) point to the
point ($\zspec,~\zspec$).  In each panel, vertical lines thus indicate
where the COMBO-17 or GOODS-MUSIC $\zphot$ is `right', while the MUSYC
$\zphot$ is `wrong'; conversely, horizontal lines show where the MUSYC
$\zphot$ is `better' than that from COMBO-17 or GOODS-MUSIC.  Note
that for the comparison to COMBO-17, we restrict our attention to
those galaxies with $R < 24$, since this is the reliability limit of
the COMBO-17 catalog.

\vspace{0.2cm}

Owing to its medium-band photometry, the COMBO-17 redshifts should be
significantly better than our own for $z \lesssim 1$, but without NIR
photometry, the redshifts of $z \gtrsim 1$ galaxies are poorly
constrained.  The agreement between the COMBO-17 photometric redshifts
and our own (left panel of Figure \ref{fig:photz}), the agreement is
indeed very good for $\zphot < 0.8$.  For $R < 24$ and $z_\mathrm{C17}
< 1.0$, the random scatter between the COMBO-17 and MUSYC photometric
redshifts is $\sigma_z = 0.034$; separately, for $R < 24$ and $\zspec
< 1$, the photometric redshift error is $\sigma_z = 0.030$ for MUSYC,
and 0.020 for COMBO-17.

There are, however, several important differences between the MUSYC
and COMBO-17 redshifts.  First, note the effect of the $z_\mathrm{C17}
< 1.4$ grid used by COMBO-17; coupled with their method of assigning
redshifts (\viz , marginalizing over the redshift probability
distribution), this means that galaxies are essentially never given
$z_\mathrm{C17} \gtrsim 1.3$.  

The exceptions to this rule are those objects that COMBO-17 has
classified as QSOs on the basis of their optical SEDs; where MUSYC
tends to place these objects at $z_\mathrm{MUS} \lesssim 1$, the
COMBO-17 redshifts are very good.  (Note that we have made no attempt
to explicitly accommodate AGNs or QSOs in our photometric redshift
calcuation.)  In the left panel of Figure \ref{fig:photz}, we mark
X-ray selected galaxies from the \citet{xray} and \citet{TreisterEtAl}
catalogs with a cross.  For this $R < 24$ sample, X-ray selected
sources are roughly three times as likely to be outliers (here, we
define outliers as those objects with $|\Delta z|/(1+\zspec) > 0.1$):
the outlier fraction for X-ray sources is 35\% (75/217), compared to
11\% (164/1438) overall.  Said another way, roughly half of all ($R <
24$) outliers are X-ray sources.


\begin{figure*} \centering
\includegraphics[width=8.8cm]{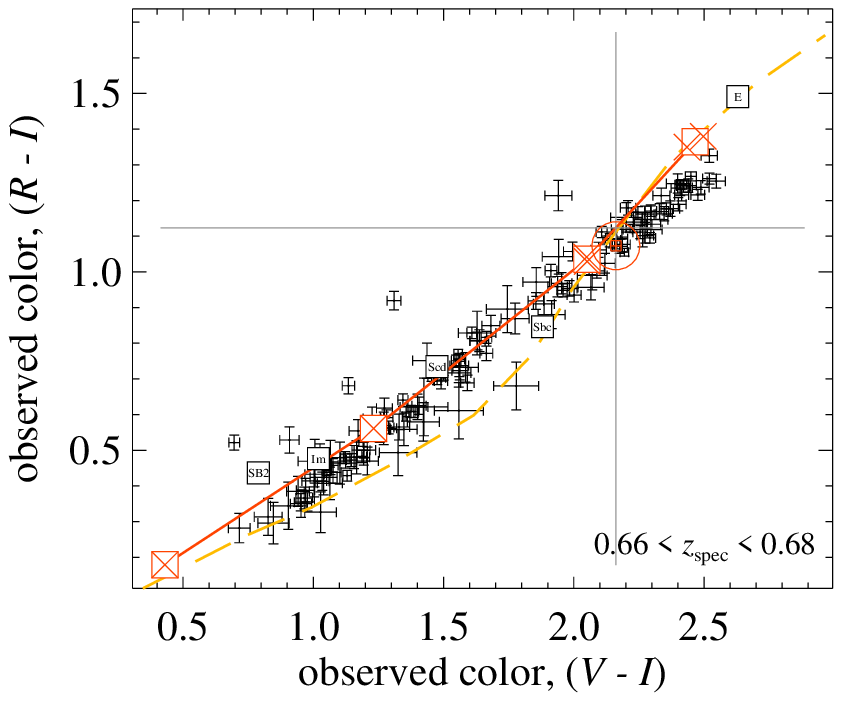}
\includegraphics[width=8.8cm]{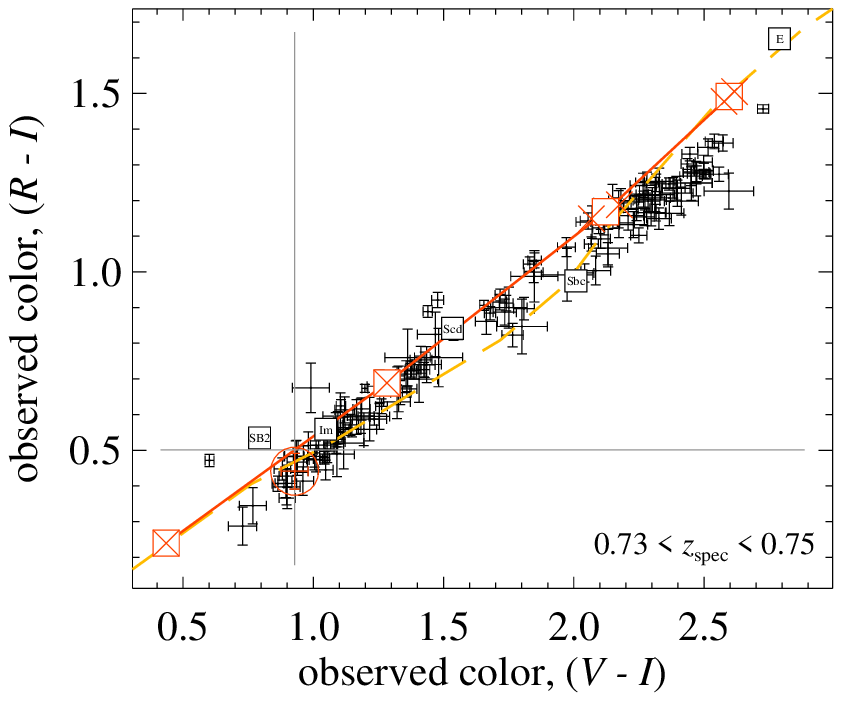}
\caption{Illustrating the InterRest algorithm for interpolating
  restframe fluxes---Note that under normal circumstances, in order to
  interpolate a restframe flux, we would relate an
  observed--minus--observed color to a restframe--minus--observed
  color; in this case we are using the $(V-I)$ color to {\em predict}
  the flux in the observers' $R$ band.  By comparing the interpolated
  and observed $R$ band fluxes, we will then be able to validate the
  algorithm (Figure \ref{fig:valid}).  The algorithm works as follows:
  using a set of template spectra ({\em red crosses}), we construct a
  (redshift-dependent) color--color relation for galaxies ({\em red
  line}); once the galaxy color--color relation has been defined, it
  is possible to read off the `unknown' color (in this case, $R-I$) of
  any object, given its known, observed color (in this case, $V-I$).
  In both panels, the points with error bars show galaxies in a narrow
  spectroscopic redshift range, with colors measured to better than
  0.05 mag; these galaxies are used in Figure \ref{fig:valid} to
  validate our restframe color determinations.  (See main text for
  further discussion and explanation.)
\label{fig:algor} }
\end{figure*}

Secondly, there are two populations of objects with $z_\mathrm{C17}
\lesssim 0.2$ that are placed by MUSYC at either $z_\mathrm{MUS} \sim
0.4$ or $z_\mathrm{MUS} \gtrsim 1.4$.  From this first population, no
$\zspec$s are available; for the second, the handful of available
$\zspec$s confirm that these galaxies are at $z \gtrsim 1.4$.  On the
other hand, for the diffuse cloud of galaxies given $z_\mathrm{MUS}
\lesssim 0.5$ and $0.5 \lesssim z_\mathrm{C17} \lesssim 1.0$, the
$\zspec$s support the COMBO-17 determinations.  

Thirdly, while objects given $0.8 \lesssim \zphot \lesssim 1.2$ in one
catalog generally lie in the same redshift interval in the other,
there is only a very weak correlation between the redshifts within
this interval: the implication here is that objects with $\zspec
\gtrsim 0.8$ are assigned $0.8 \lesssim z_\mathrm{C17} \lesssim 1.2$
more or less at random on the basis of optical data alone.  In other
words, while the COMBO-17 $\zphot$---$\zspec$ agreement is excellent
for $\zspec \lesssim 0.8$, a $0.8 \lesssim z_\mathrm{C17} \lesssim
1.0$ selected sample may suffer significant contamination from $\zspec
\gtrsim 1$ galaxies with poorly constrained redshifts.


\vspace{0.2cm}

Looking now at the comparison with the GOODS-MUSIC redshifts (right
panel of Figure \ref{fig:photz}), it is clear that, while the random
scatter between the two determinations is larger than for the previous
comparison, at least for $z \lesssim 1$, there are no signs of major
systematic discrepancies.  The random scatter between the GOODS-MUSIC
and MUSYC photometric redshifts is $\sigma_z = 0.065$; separately, for
the same $\zspec$ comparison sample, the random errors are $\sigma_z =
0.036$ for MUSYC, and 0.043 for GOODS-MUSIC.  Both MUSYC and
GOODS-MUSIC suffer from catastrophic failures, where $\zspec \sim 0.7$
galaxies are given $\zphot \sim 0.2$; although this appears to be a
greater problem for MUSYC.  GOODS-MUSIC also seems to have some
systematic issues for $\zphot \approx 0.4$.

In this panel, we mark with a circle those objects with $Q_z > 3$.
Whereas roughly half (938/1787) of the objects plotted in this panel
have robust $\zspec$s, the fraction among those with $Q_z > 3$ is just
33\% (242/735); again, this underscores the importance of having a
representative spectroscopic comparison sample.  Using the cut
$|(z_\mathrm{MUS}-z_\mathrm{GDS}|/(1+z_\mathrm{GDS}) > 0.2$ to
quantify the level of disagreement between the GOODS-MUSIC and MUSYC
redshifts, objects with $Q_z > 3$ are twice as likely to be outliers:
the fraction is 60\% (99/166) for $Q_z > 3$, compared to 33\%
(586/1787) overall.  We note that the fraction of sources with $Q_z >
3$ increases from $\lesssim 5$\% for $z_\mathrm{GDS} \lesssim 1.2$ to
$\sim 15$ \% for $1.2 \lesssim z_\mathrm{GDS} \lesssim 2.2$.  For
$z_\mathrm{GDS} > 2.5$, roughly half (9/21) of all galaxies have $Q_z
> 3$.  Similarly, X-ray-selected galaxies are more likely to be
outliers: the outlier fraction for X-ray sources is 43\% (16/37).

Again, we caution that, without spectra for a large, representative
subsample of the objects common to these two catalogs, it is not
possible to determine whether one catalog is truly `better' than the
other.  Moreover, given the differences between the MUSYC and
GOODS-MUSIC catalogs---particularly the inclusion of ACS and IRAC
imaging in the GOODS-MUSIC catalog---it is not possible to say whether
any differences in photometric redshifts are due to the photometric
redshift algorithms or to differences in the data themselves.  Given
these differences, however, the broad agreement between the MUSYC and
GOODS-MUSIC $\zphot$s, and especially for $\zphot \gtrsim 1$ where
$\zspec$s are increasingly hard to come by, is certainly encouraging.


\section{Interpolating Restframe Photometry --- Introducing I\lowercase{nter}R\lowercase{est}}
\label{ch:interrest}

Given an SED and a redshift, we have derived restframe photometry
following the method described in Appendix C of
\citet{RudnickEtAl2003}.  This method is best understood as
interpolating between two points in the observed SED to come up with a
restframe flux.  We have developed an IDL implementation of this
algorithm for interpolating restframe photometry, dubbed InterRest.
InterRest has been specifically designed to dovetail with EAZY: it
accepts the same inputs and configuration files, and uses the same
algorithms for integration, etc.  We have made this utility freely
available to the astronomical community.

The essential idea is to use a set of template spectra to construct a
color--color relation for galaxies at a given redshift.  Specifically,
we relate a color in terms of two observed filters to another color in
terms of an observed filter and the desired restframe filter.  For
example, in order to find the restframe $r$ flux of a galaxy at $z =
1.2$ ($\lambda_\mathrm{em} = 6220$ \AA; $\lambda_\mathrm{ob} = 13700$
\AA), we would relate the ($z' - J$) color to the ($r_{z=1.2}-J$)
color; the $r_{z=1.2}$ flux then immediately follows.

This process is illustrated in Figure \ref{fig:algor}, with one
crucial difference: whereas normally, in order to interpolate a
restframe flux, we would relate an observed--minus--observed color to
a restframe--minus--observed color, in this example we are concerned
with using the observed $(V-I)$ color to {\em predict} the observed
$(R-I)$ color, and so the observed $R$ flux.  In this way, we will be
able to test the accuracy of the algorithm, through comparison between
the predicted and observed $R$ fluxes.  Even so, the example still
serves to illustrate the idea behind the algorithm.

In each panel of Figure \ref{fig:algor}, the points show the observed
$VRI$ colors of galaxies with spectroscopic redshifts in a narrow
interval; we have selected the two most prominent redshift spikes, and
restrict our attention to galaxies with colors measured to better than
0.05 mag.  The red crosses in each panel show the synthetic $VRI$
colors for the default EAZY/InterRest template set, which we use to
construct an approximate color--color relation for galaxies at each
redshift.  In both panels, the default EAZY/InterRest template spectra
can be seen to do a reasonable job of describing the true color--color
relation for galaxies at each of the two redshifts in question.

Now, for any individual galaxy (red point, circled; chosen at random),
using the $(V-I)$ color, it is possible to read off the $(R-I)$ color
(grey lines) from the synthetic color--color relation..  Again, under
normal circumstances, we would be relating an
observed--minus--observed color to an restframe--minus--observed
color; our interest here is in validating the performance of the
algorithm.

As a single algorithmic detail, it is possible that the known--known
colors (\ie , ($V-I$) in the above example) of two templates are very
close, but for quite different known--unknown colors (\ie , ($R-I$)
above): in this case, small changes in color or redshift can produce
very large changes in the final result.  To avoid this situation,
where the known--known colors are too close, we simply replace these
points with their mean (in magnitude space).  This can be seen in
Figure \ref{fig:algor}, where the crosses show the points for the
individual template spectra, and the squares show the points used to
construct the color--color relation.  Algorithmically, we define `too
close' as two points being separated by less than 5 \% of the range
spanned by all template spectra.


\begin{figure*} \centering
\includegraphics[width=15cm]{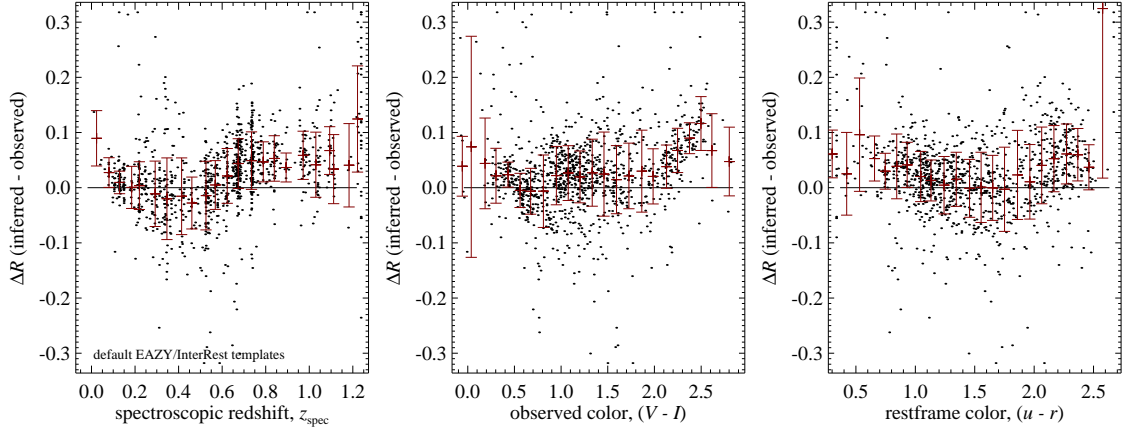}
\caption{Validating the InterRest algorithm for interpolating
  rest-frame photometry---In each panel, we show the difference,
  $\Delta R$, between the $R$ band magnitude inferred from the $(V-I)$
  color as in Figure \ref{fig:algor} and that directly observed,
  plotted as a function of ({\em left to right}) spectroscopic
  redshift, observed color, and rest-frame color.  The points in each
  panel show galaxies with robust spectroscopic redshift
  determinations, and $VRI$ colors measured to better than 0.05 mag;
  the red points with error bars show the biweight mean and scatter in
  $\Delta R$.  The random error in the interpolated $R$ band magnitude
  is typically $\lesssim 0.05$ mag; comparable to the observational
  uncertainties themselves.  Systematic uncertainties, as functions of
  both redshift (\ie\ restframe wavelength) and restframe color are at
  the level of $\lesssim 0.05$ mag.  Note, however, that the
  (logarithmic) wavelength interval between $V$ and $I$ is roughly
  twice as large as we would normally use to derive rest-frame
  photometry for real galaxies.  We therefore present these numbers as
  upper limits on the true errors; we expect the true errors to be
  smaller by a factor of 2---4.
\label{fig:valid} }
\end{figure*}

In Figure \ref{fig:valid}, we show the differences between the $R$
fluxes interpolated as described above, and the observed $R$ fluxes in
the MUSYC catalog, plotted as a function of ({\em left to right}),
spectroscopic redshift, observed color, and restframe color.  These
plots are based on the $\zspec$ compilation used in Figure
\ref{fig:photz}, and described in Appendix \ref{ch:speczs}, but
limited to those galaxies with $VRI$ colors measured to better than
0.05 mag.  The black points are for individual galaxies; the red error
bars show the mean error and random scatter in bins.

\begin{table*} \centering
\caption{Summary of the Contents of the Restframe Photometry Catalog}
\begin{tabular*}{0.95\textwidth}{@{\extracolsep{\fill}}l l p{12 cm}}
\hline\hline \label{tab:rfphotcat}
\\
Column No. & Column Title & Description \\
\\
\hline\hline
\\

1 & id & Object identifier, beginning from 1, as in the photometric catalog \\

2 & redshift & Assumed redshift; we use either the z\_m2 value output
by EAZY, or the spectroscopic redshift, where available. \\

3---17 & RF\_F1, etc. & Restframe photometry for Bessel
$UBVRI$ filters$^a$ \\

18---32 & RF\_F6, etc. & Restframe photometry for Johnson--Cousins $UBVRI$
filters \\

33---47 & RF\_F11, etc. & Restframe photometry for Gunn $ugriz$ filters \\

48---54 & RF\_F16, etc. & Restframe photometry for GALEX NUV and FUV filters \\

55 & distmod & The distance modulus implied by redshift, assuming a
given cosmology$^b$ \\
\\
\hline
\hline
\end{tabular*}
\tablecomments{$^a$ For each object and filter, InterRest outputs two
  flags: extrap$n$ (where $n$ refers to the restframe filter number),
  which indicates where it has extrapolated beyond the observed SED,
  and widegap$n$, which indicates where it has not used neighboring
  filters due to, for example, missing or negative photometry. $^b$
  Note that the fluxes output by InterRest are {\em observed fluxes}
  through {\em restframe filters}; that is, they have the same units
  as the observed, input photometry.  The user must therefore perform
  the conversion to apparent and restframe magnitudes using the
  appropriate zeropoint and distance modulus.}
\end{table*}

Both the random scatter and the systematic offset between the observed
and interpolated $R$ fluxes are at the level of 0.05 mag.  There are
clear systematics with redshift (\ie\ restframe wavelength), which
appear to be related to the 4000 \AA\ break.  There also appears to be
a problem at the level of 0.05 mag for the reddest galaxies ($u-r
\gtrsim 2$).  The random error in the interpolated $R$ fluxes is
typically $\sim 0.05$ mag.  This is comparable to the uncertainties
in the photometry itself, but probably at least partially reflects the
intrinsic width of the galaxy color--color(redshift) sequence; if so,
this represents a fundamental limit on the accuracy of the algorithm.

Note that whereas we would typically use two neighboring filters to
interpolate restframe photometry, the wavelength span here is roughly
twice as large; we therefore expect the true systematic errors in
restframe fluxes (\cf\ colors) to be 2---4 times smaller than in the
above example; \ie\ at the level of 0.01---0.02 mag.

As a final aside, we note that we acheive comparable accuracies using
the E, Scd, Sbc, and Im templates from \citet{cww}, supplemented with
a starburst template from \citet{Kinney}.  These templates are plotted
in Figure \ref{fig:algor} (black squares, labeled) for comparison to
the default EAZY/InterRest templates.  We have also tried using
\citet{BruzualCharlot} synthetic spectra, assuming Single Stellar
Populations (SSPs; $\log t$ = 6.5, 7.0, ..., 10.0, 10.3 Gyr) with a
Salpeter IMF and solar metallicity, and no dust extinction (shown by
the dashed yellow line in Figure \ref{fig:algor}).  Using BC03
spectra, we find serious systematic errors --- on the level of up to
0.2 mag --- both as a function of redshift, and of restframe color;
this is true whether we assume a SSP or exponentially declining star
formation history.  These models do not reproduce the observed colors
of real galaxies, and so are unsuitable for this purpose.  Similarly,
using the \citet{blanton:photz} template set, which are derived from a
library of BC03 spectra with a wide range of ages and metallicities
using the NMF algorithm, we find peak--to--peak systematic errors at
the $\sim 0.1$ mag level; the random errors are also at the 0.1 mag
level.

In Table \ref{tab:rfphotcat}, we summarize the contents of the
restframe photometry catalogs that we are releasing: note that we
provide two separate catalogs based on photometric and spectroscopic
redshift determinations, respectively.


\section{Summary} \label{ch:summary}

We have described a new $K$--selected catalog of the ECDFS based on
existing optical and NIR data, supplemented by original $z'JK$ imaging
taken as part of the MUSYC project.  The final $UU_{38}BVRIz'JHK$
photometric catalog (\textsection\ref{ch:catalog}; Table
\ref{tab:photcat}) covers $\sim 900~ \square~''$ to a ($5\sigma$,
point source) limiting magnitude of $K = 22.0$ mag; note, however,
that $H$ band data is available for only 80 \% of the field.  Included
in the photometric catalog are a spectroscopic redshifts for 2914
unique objects, collected from the literature (Appendix
\ref{ch:speczs}).  In addition, we are also making available a
photometric redshift catalog, derived from the MUSYC ECDFS photometry
using EAZY (\textsection\ref{ch:eazy}; Table \ref{tab:photzcat}), as
well as catalogs of interpolated restframe photometry generated using
InterRest (\textsection\ref{ch:interrest}; Table \ref{tab:rfphotcat}).

The data described in this paper will form an important part of two
ongoing NIR survey projects.  The $K$ imaging is key for analysing the
SIMPLE IRAC data (Damen et al., in prep.).  The broadband imaging
provides the backbone for an optical medium-band survey, which will
add 18 additional bands (Cardamone et al., in prep.).  There is also a
NEWFIRM medium band NIR survey planned, which will allow much greater
photometric redshift accuracy for $z \gtrsim 1$ \citep{newfirm}.  We
have invested significant time and effort in validating the absolute
and relative calibration of the imaging data, as well as our analysis
techniques, so as to maximize the legacy value of our catalogs.  We
summarize the results of these checks below.

{\em Astrometry}--- The relative astrometric calibration of each band
has been validated to $0\farcs15$ (0.56 pix).  In absolute terms, the
absolute astrometry is accurate to $0\farcs3$ (1.12 pix), with a
slight shear across the field at the level of $0\farcs1$ (0.37 pix;
see \textsection\ref{ch:astromcomps}).

{\em Completeness}---We have quantified the completeness of the
catalog for sources with an $R^{1/4}$ profile in Figure
\ref{fig:completeness}; we present these values as lower limits on the
completeness.  While the catalog is formally surface brightness
limited, a comparison to much deeper NIR imaging over the GOODS area
of the field suggests that the catalog is more nearly flux limited.
This comparison suggests that for $K = 22$, the catalog is $\sim$
85---90 complete, and $\gtrsim 95$ \% reliable
(\textsection\ref{ch:completeness}).

{\em Photometric Calibration}---While there are significant
differences between the photometry in the COMBO-17 and MUSYC
catalogs of the ECDFS (\textsection\ref{ch:combo}), a comparison
between the MUSYC and GOODS photometry in the region of overlap
validates the MUSYC photometry to $\lesssim 0.05$ mag
(\textsection\ref{ch:goods}).  We have refined the basic photometric
calibration using the observed SEDs of main sequence stars; we
estimate that after this recalibration, the photometric
cross-calibration is accurate to $\lesssim 0.02$ mag
(\textsection\ref{ch:starcols}).

{\em Photometry}---Random and systematic photometric errors due to
various aperture effects (including astrometric errors and imperfect
PSF matching) are limited to $\lesssim 0.03$ mag and $\lesssim 0.006$
mag, respectively (Figure \ref{fig:psfmatching}).  We have applied
corrections to SExtractor's AUTO flux measurements to account for
missed flux and background oversubtraction; for synthetic
$R^{1/4}$--law sources, these corrections typically reduce the offset
between the known and recovered total fluxes by 0.05---0.10 mag
(\textsection\ref{ch:totalmags}).  We have also demonstrated that the
photometric errors given in the catalog accurately trace variations
in the background RMS in the NIR images (Figure \ref{fig:errors}).

{\em Spectroscopic Redshifts}---We have collected and collated 5374
spectroscopic redshift determinations from literature sources, of
which 3815 are matched to 2914 unique sources in our catalog
(Appendix \ref{ch:speczs}).  Of these, 2213 redshifts are deigned
`secure', including 247 stars, and 1966 $z \gg 0$ galaxies.

{\em Photometric Redshifts}---There are some systematic discrepancies
between the COMBO-17 and MUSYC photometric redshift determinations in
the ECDFS, owing to the lack of NIR data in the COMBO-17 catalog;
where available, spectroscopic redshifts validate the MUSYC values.
The agreement between the MUSYC and GOODS-MUSIC photometric redshifts
is very good, however there are a significant number of catastrophic
errors in both redshift catalogs (Figure \ref{fig:photz}).  In
comparison to spectroscopic redshifts from the K20 survey
\citep{K20survey, k20specz}, the random photometric redshift error is
$\sigma_z = 0.033$, with an outlier fraction of 4.7 \%; the outlier
fraction is significantly higher for X-ray--selected spectroscopic
redshift catalogs (Appendix \ref{ch:speczs}).

{\em Restframe Colors}---We have interpolated restframe photometry for
the galaxies in our catalog using an IDL utility called InterRest
(\textsection\ref{ch:interrest}); we also make this utility publicly
available.  Estimated systematic errors in these interpolated
restframe fluxes, as functions both of restframe wavelength and of
galaxy color, are estimated to be $\lesssim 0.02$ mag (Figure
\ref{fig:valid}).  Random errors inherent to the algorithm are at a
similar level.

\vspace{0.2cm}

The primary science application of the $K$--selected catalog that we
have presented here is to characterise the properties of massive
galaxies at $z \lesssim 2$, including their evolution.  In Paper II,
we demonstrate that this catalog is approximately complete (volume
limited) for $M_* \gtrsim 10^{11}$ M\sun\ and $\zphot \lesssim 1.8$,
and use this catalog to quantify the $z \lesssim 2$ evolution in
number density and color of massive galaxies in general, and of red
sequence galaxies in particular.

In this context, the MUSYC ECDFS dataset provides a valuable
complement to existing optical surveys in the ECDFS targeting the $z
\lesssim 1$; \eg\ the COMBO-17 \citep{WolfEtAl} and GEMS projects
\citep{RixEtAl}.  Further, the $z \lesssim 2$ comoving volume
contained with the ECDFS field is approximately three times greater
than that at $z \lesssim 3.5$ within the GOODS region in the CDFS.
The MUSYC ECDFS catalog thus also complements the much deeper
GOODS-CDFS data, by allowing better sampling of rare objects,
including the most massive galaxies at moderate-- to high--redshifts.
Taken together, these combined datasets form an outstanding laboratory
to study the basic properties of galaxies over nearly 90 \% of the
history of the universe.

This work was supported through grants by the Nederlandse Organisatie
voor Wetenschappelijk Onderzoek (NWO), the Leids Kerkhoven-Bosscha
Fonds (LKBF), and National Science Foundation (NSF) CAREER grant AST
04-49678.  We thank the referee, Stefano Berta, for a close and
thorough reading of the manuscript, which helped to clarify a number
of points.  We also wish to thank the organizers and participants of
the several workshops hosted by the Lorentz Center, where many aspects
of this work were developed and refined.  SW gratefully acknowledges
support from the W M Keck Foundation.



\appendix


\section{A Compilation of Public Spectroscopic Redshift Determinations for the MUSYC ECDFS Catalog} \label{ch:speczs}

\begin{table*}\begin{center}
\caption{Summary of the Spectroscopic Redshifts Available for MUSYC
ECDFS detections}
\begin{tabular*}{0.95\textwidth}{@{\extracolsep{\fill}}l c c c c c c c}
\hline
\hline
\multicolumn{1}{c}{Reference(s)}
& Source & Internal & No. & No. & Median & NMAD & Outlier \\
& Code & Qual.\ Flag & Galaxies & Adopted & $\Delta z/(1+z)$ & $\Delta z / (1+z)$ & Fraction \\
\multicolumn{1}{c}{(1)} & (2) & (3) & (4) & (5) & (6) & (7) & (8) \\
\hline
\citet{K20survey};  & K20&1&         267&         232&  -0.025&   0.033&   0.047\\
\hspace{0.2cm} \citet{k20specz} & &0&  14&        2&  -0.012&   0.069&   0.182\\
\citet{xray}  & Xray&$\ge 2.0$ &   114&         114&  -0.024&   0.037&   0.135\\
              &     &$<2.0$&        17&           4&   0.045&   0.146&   0.133\\
\citet{vvds} &     VVDS&4&         172&         131&  -0.030&   0.027&   0.027\\
                 &     &3&         347&         267&  -0.030&   0.032&   0.035\\
                 &     &2&         342&          19&  -0.022&   0.058&   0.080\\
                 &     &1&          82&           1&  -0.003&   0.127&   0.017\\
                 &     &9&          49&           1&   0.016&   0.199&   0.036\\
\citet{fors21, fors22, fors23} & GDS-F&A& 306&  226&  -0.023&   0.044&   0.034\\
                &      &B&          77&          14&  -0.029&   0.080&   0.054\\
                &      &C&          52&           4&   0.025&   0.106&   0.079\\
\citet{vimos}   & GDS-V&A&         289&         197&  -0.036&   0.030&   0.048\\
                &      &B&          59&           3&  -0.026&   0.081&   0.087\\
                &      &C&          48&           1&  -0.008&   0.144&   0.051\\
\citet{images} & IMAGES&1&         267&         219&  -0.032&   0.030&   0.067\\
               &       &2&         168&          24&  -0.025&   0.046&   0.056\\
               &       &3&          51&           7&  -0.012&   0.095&   0.000\\
\citet{TreisterEtAl} & MUS-I&N/A&   165&         120&   0.001&   0.112&   0.125\\
              & MUS-V&N/A&          34&          33&   0.011&   0.295&   0.000\\
\citet{KoposovEtAl} & Kopsv& N/A & 455&         283&  -0.034&   0.025&   0.043\\
\\
\citet{kx}      & KX&N/A &          17&           5&  -0.016&   0.029&   0.353\\
\citet{sne}    & SNe&N/A &           9&           2&  ---   &   ---  &   ---  \\
\citet{vdwel04,vdwel05} & vdWel &N/A&  28&       26&  -0.007&   0.022&   0.000\\
\citet{Daddi2005} & Daddi&N/A&       5&           5&  ---   &   ---  &   ---  \\
\citet{lcirs}  & LCIRS&1---3&       14&          10&   0.003&   0.050&   0.071\\
\citet{kriek}   & Kriek&N/A&        12&          12&   0.056&   0.134&   0.000\\
\hline
\\
 Total & &  &        2863 &         1966&  -0.029&   0.036&   0.078\\
\\
\hline
\hline
\end{tabular*} \end{center}
\tablecomments{For each spectroscopic redshift sample we have used, we
  give both the redshift source catalog (1) and the identifier used in
  the MUSYC $\zspec$ catalog (2); further, we have broken up each
  sample by the internal quality flag (3), where available.  For each
  (sub)sample, we give the number of galaxies matched to the MUSYC
  ECDFS catalog (4), and the number of galaxy redshifts adopted in the
  final catalog (5).  We also give the systematic (6) and random (7)
  photometric redshift error, computed as the median and NMAD of
  $\Delta z/(1+z)$, and the outlier fraction (8), defined as the
  fraction of galaxies with $\Delta z/(1+z) > 0.1$; these quantities
  are all computed for galaxies in our main scientific sample (\ie\
  those galaxies counted in column 4 with coverage in optical and NIR
  bands, and with $K < 22$ and $K$ S:N $> 5$).
\label{tab:speczs}}
\end{table*}

The ECDFS has been targeted by a number of large spectroscopic
redshift campaigns, including: optical spectroscopy of the original
CDFS X-ray catalog by \citet{xray}, the K20 survey \citep{K20survey,
k20specz}, the VVDS \citep{vvds}, the GOODS FORS2 \citep{fors21,
fors22, fors23} and VIMOS \citep{vimos} campaigns, the IMAGES survey
\citep{images}, a MUSYC program targeting X-ray sources in the full
ECDFS \citep{TreisterEtAl}, and a VIMOS campaign by
\citet{KoposovEtAl}.  A summary of the spectroscopic redshift
resources we have used is given in Table \ref{tab:speczs}.
Altogether, we have collected 5374 separate spectroscopic redshift
determinations, of which 3815 are matched to 2914 unique objects in
our catalog.

In cases where multiple spectroscopic redshift
determinations/identifications are available for individual objects,
our guiding principles for selecting a redshift were as follows.
First, we adopt the most common redshift determination (where $\Delta
z < 0.01$ is taken as agreement, and we do not consider repeat
observations by the same team as an independent measurement).  574
objects in the catalog have multiple, consistent redshift
determinations.  Where there is no consensus, we discriminate between
redshift solutions on the basis of the $Q_z$ figure of merit developed
by \citet{eazy}, evaluated for the spectroscopic redshift.  An
exception to this rule is for redshifts from the X-ray selected
catalogs, which do occasionally have extremely high values of
$Q_z(\zspec)$, even when confirmed by other secure determinations from
other catalogs.  Where $Q_z(\zspec)$ does not clearly discriminate
between the possible solutions, we fall back onto the quality flags
given by the different spectroscopic surveys.  Note that for this
purpose, we do not consider the VVDS `2' flag as `secure'.  Similarly,
we give preference to the results of smaller studies, which presumably
have devoted greater care on a per object basis.  Reassuringly, in
almost all cases, these criteria reinforce one another.  Finally, we
choose to adopt redshifts from sources that provide classification
information where available; this means that we tend not to adopt
redshifts from, for example, the VVDS catalog where other
determinations are available.  Moreover, we consider X-ray selection
as an additional piece of classification information; accordingly, we
adopt redshifts from the \citet{xray} and \citet{TreisterEtAl}
catalogs where available.

In this way, we have constructed a compendium of spectroscopic redshift
determinations for 2914 unique objects in the MUSYC ECDFS catalog,
including 283 spectrally-classified stars.  Although all of these
determinations are given in the catalog, we will only consider those
deigned `secure', either by virtue of their quality flags, or through
agreement between multiple sources.  This leaves 2213 robust
spectroscopic redshifts for objects in the MUSYC catalog; 1966 of
these objects are identified as $z \gg 0$ galaxies.


\begin{figure*} \centering
\includegraphics[width=15.5cm]{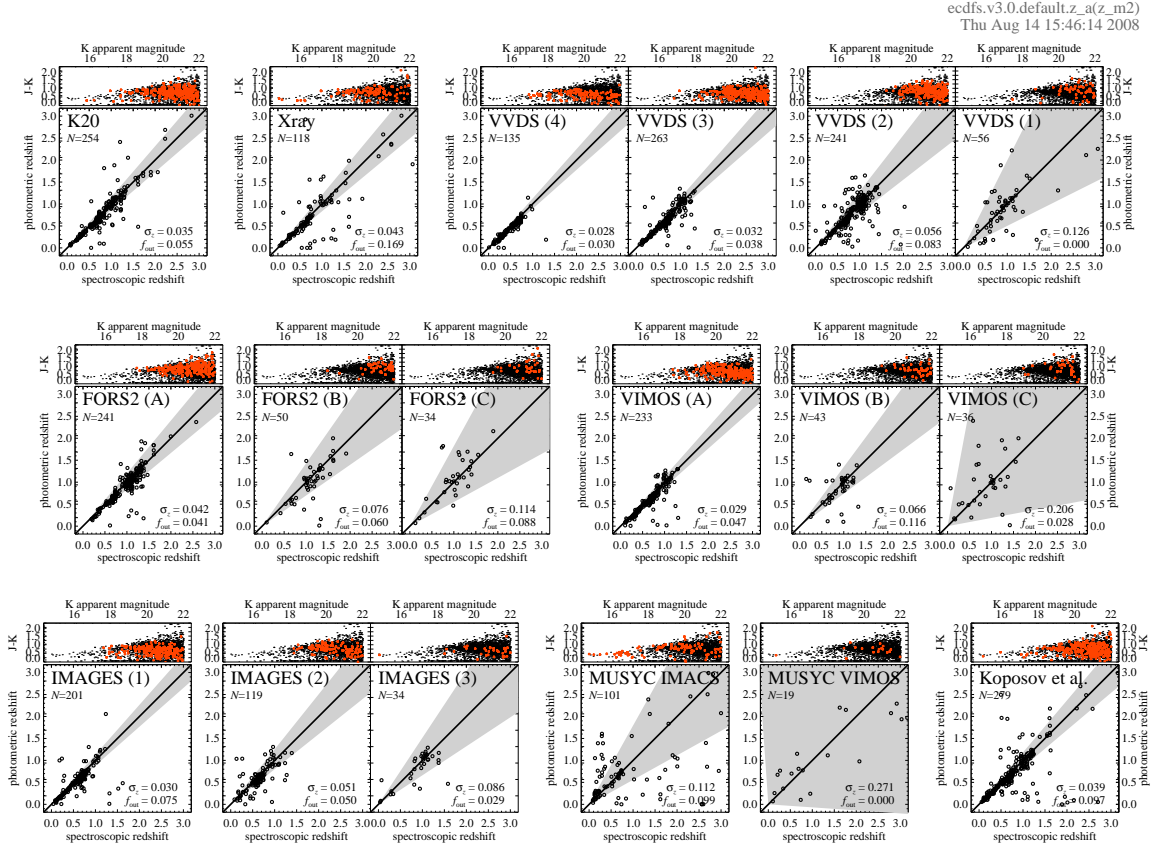}
\caption{The $\zphot$---$\zspec$ diagram for individual spectroscopic
  redshift sources and quality flags---We show the agreement between
  the MUSYC photometric redshift and many literature spectroscopic
  redshift catalogs, broken up by internal quality flag where
  available.  Within each panel, we give the random photometric
  redshift error, $\sigma_z$, as well as the outlier fraction, which
  is defined as the fraction of objects with $\Delta z/(1+z) > 5
  \sigma_z$.  The grey regions in each panel indicate the 3$\sigma_z$
  regions.  Above each $\zphot$---$\zspec$ diagram, we highlight the
  particular $\zspec$ sample in $(J-K)$---$K$ color--magnitude space.
  It is harder to obtain robust $\zspec$ determinations for redder
  galaxies; these galaxies are therefore underrepresented in all
  $\zspec$ samples.  For this reason, we have validated our
  photometric redshift determinations through comparison with those
  from COMBO-17 and GOODS-MUSIC (see Figure \ref{fig:photz}).
\label{fig:photz-specz} } \end{figure*}

Figure \ref{fig:photz-specz} shows the $\zphot$---$\zspec$ diagram,
broken up by the $\zspec$ source catalog, and quality flag.  Within
each panel, we give the NMAD and median offset in $\Delta z / (1+z)$;
these values are also given in Table \ref{tab:speczs}; the grey region
indicates the 3$\sigma_z$ errors around the $\zphot = \zspec$ line.
Above each $\zphot$---$\zspec$ diagram, we also show the distribution
of each $\zspec$ sample in observed $(J-K)$---$K$ color---magnitude
space, in comparison to the full MUSYC catalog.

For `secure' redshift determinations, the $\zphot$--$\zspec$ agreement
is really quite good: the typical random scatter is $\sigma_z \lesssim
0.040$.  Particularly for $\zspec \lesssim 1$, we do appear to
slightly underestimate galaxies' redshifts; typical systematic errors
are $\Delta z/(1+z) \sim -0.025$.  For the \citet{xray} catalog, the
random scatter in $\zphot$ determinations is still quite good, but for
the MUSYC spectroscopic redshift program \citep{TreisterEtAl}, which
targets brighter X-ray sources, the $\zphot$---$\zspec$ agreement is
poor.

Further, while the outlier fraction is generally at the level of a few
percent, catastrophic redshift failures appear to be a significant
problem for X-ray selected sources.  (Recall that we make no attempt
to explicitly incorporate AGNs or QSOs in our photometric redshift
calculation.)  Among X-ray-selected sources, the fraction of galaxies
with $|\Delta z|/(1+z) > 0.15$ is 30\% (82/271); for the full gamut of
robust spectroscopic redshifts, the fraction is 9\% (178/1966).  Said
another way, 46 \% (82/178) of all outliers are X-ray sources.

We draw particular attention to the comparison with the results from
K20, which is highly spectrally complete in the magnitude range that
we are operating in.  In comparison to the K20 redshifts, we have
achieved a photometric redshift accuracy of $\sigma_z = 0.034$.  We
also draw attention to the sample of \citet{vdwel05}, which consists
of 28 early type, red sequence galaxies at $z \sim 1$, for which we
have achieved a photometric redshift accuracy of $\sigma_z = 0.022$;
in fact, this is the sample for which we have the best photometric
redshift agreement.

The crucial point to be made from Figure \ref{fig:photz-specz},
however, is that since most of the different $\zspec$ samples that are
available in the ECDFS are not NIR--selected, they are not generally
representative of the sources in our photometric catalog.  For this
reason, we validate our photometric redshift determinations in
\textsection\ref{ch:photz-valid} through comparison with the COMBO-17
and GOODS-MUSIC photometric redshifts.

\end{document}